\documentclass[useAMS,usenatbib]{mnras}

\usepackage{amssymb,amsmath,amsfonts,graphicx}
\usepackage{epsfig,color} 
\usepackage[T1]{fontenc}
\usepackage{aecompl}
\usepackage{times}
\newcommand{\ud}{\mathrm{d}}

\newcommand\avg[1]{{\langle #1 \rangle}}

\def\lsim{ \lower .75ex \hbox{$\sim$} \llap{\raise .27ex \hbox{$<$}} }

\title[]{What to expect from dynamical modelling of galactic haloes
}

\author[Wang et al.]{Wenting Wang$^{1,2}$, Jiaxin Han$^{1,2}$\thanks{jiaxin.han@ipmu.jp}, Shaun Cole$^{1}$, Carlos Frenk$^{1}$, Till Sawala$^{1}$ \\
  {}$^{1}$Institute for Computational Cosmology, University of Durham, South Road, Durham, DH1 3LE, UK\\
  $^2$ Kavli IPMU (WPI), UTIAS, The University of Tokyo, Kashiwa, Chiba 277-8583, Japan
}

\begin{document}



\maketitle

\begin{abstract}

Many dynamical models of the Milky Way halo require assumptions that the distribution function 
of a tracer population should be independent of time (i.e., a steady state distribution function) 
and that the underlying potential is spherical. We study the limitations of such modelling 
by applying a general dynamical model with minimal assumptions to a large sample of galactic 
haloes from cosmological $N$-body and hydrodynamical simulations. Using dark matter particles 
as dynamical tracers, we find that the systematic uncertainties in the measured mass and 
concentration parameters typically have an amplitude of 25\% to 40\%. When stars are used 
as tracers, however, the systematic uncertainties can be as large as a factor of $2-3$. The 
systematic uncertainties are \emph{not reduced} by increasing the tracer sample size and vary 
stochastically from halo to halo. These systematic uncertainties are mostly driven by 
underestimated statistical noise caused by correlated phase-space structures that violate 
the steady state assumption. The number of independent phase-space structures inferred from 
the uncertainty level sets a limiting sample size beyond which a further increase no longer 
significantly improves the accuracy of dynamical inferences.  The systematic uncertainty 
level is determined by the halo merger history, the shape and environment of the halo. Our 
conclusions apply generally to any spherical steady-state model.

\end{abstract}


\section{Introduction}
\label{sec:intro}
Since dark matter does not emit or absorb electromagnetic radiation, 
gravitational modelling is essential to study its distribution in 
the Universe. For large samples of distant galaxies, gravitational 
lensing is the most efficient way to measure the underlying mass 
distribution \citep[e.g.][]{2001PhR...340..291B,Mandelbaum2006a,
2010MNRAS.404..486H,2015MNRAS.446.1356H}. Combining gravitational 
lensing with the dynamical modelling of Integral Field Spectrograph 
(IFU) data and stellar population synthesis modelling, the baryonic 
mass and dark matter distributions can be modelled and constrained 
simultaneously for relatively bright  galaxies and over a wide range 
of radius \citep[e.g.][]{2012Natur.484..485C,2015MNRAS.446..493P}. 

Compared with these more distant galaxies, our Milky Way (MW) is special.
Because of the closeness, the dynamical information of individual MW halo 
stars can be resolved. Moreover, since we are embedded in the MW, both 
the radial velocities along the line-of-slight and tangential velocities 
perpendicular to the radial direction can be observed. Thus, there are 
many different methods that can be applied  to infer the mass distribution 
in the MW using bright halo stars, satellite galaxies and globular clusters 
as dynamical tracers \citep[e.g.][]{2007MNRAS.379..755S,2011ApJ...743...40B, 
2013ApJ...770...96G,2013ApJ...768..140B,2013ApJ...773L..32R,2014MNRAS.437..959B,
2014A&A...562A..91P}. A more detailed summary of these methods of measuring 
the MW halo mass is available in \cite{2015MNRAS.453..377W}.  

These methods depend on various assumptions. For example, most of the models 
assume tracers are in a steady state and the underlying potential is spherical. 
To constrain the mass profile using the Jeans equation, additional assumptions about 
the velocity anisotropy of tracers usually have to be made. At least partly 
due to violations of all these model assumptions, the dynamically inferred masses 
of the MW in the literature have large uncertainties and cover a wide 
range from $0.5$ to $2.5\times 10^{12}~\mathrm{M_\odot}$ \citep[e.g.][]{1999MNRAS.310..645W,
2005MNRAS.364..433B,2008ApJ...684.1143X,2008MNRAS.384.1459L,2012MNRAS.424L..44D,
2014MNRAS.445.3788G,2014MNRAS.443.2204P,2015ApJ...806...54E}. The MW halo mass and the 
inferred satellite galaxy properties, however, play a crucial role in many inferences derived 
from the properties of the MW or Local Group system \citep[e.g.][]{2011MNRAS.415L..40B,
2012MNRAS.424.2715W,2012MNRAS.424...80P,Cautun_2014a,2015MNRAS.449.2576C,2015MNRAS.452.3838C}. 
Thus more accurate measurements are necessary for robust cosmological inferences, 
which requires proper understanding of these model assumptions.

In a previous paper \citep{han2015a}, we developed the orbital 
Probability Distribution Function (oPDF) method to infer the underlying 
mass distribution or halo potential, which only assumes that tracers are 
well phase mixed and relaxed, and the underlying potential is spherical. 
This clean method allowed us to further investigate how the two 
assumptions hold when it is applied to Aquarius haloes \citep{han2015b}. 
We found dark matter particles are more relaxed than stars. 
Using dark matter particles as tracers, we can reach an accuracy of 
5\% in the inferred halo mass if the underlying potential profile is 
modelled properly, whereas there is about a 20 to 40\% systematic 
uncertainty when using stars as tracers. The outcome tells us the accuracy 
to expect from the dynamical modelling of MW halo stars. These results 
are also relevant to other methods which also make the steady state and 
spherical assumptions as our results are expected to have the minimum 
systematic uncertainty one can achieve under these two model 
assumptions.

This conclusion, however, is based on the limited statistical power of
only five haloes.  In this paper, we extend the analysis by applying 
the oPDF model to a much larger sample of ($\sim 1000$) MW size haloes 
selected from the Millennium~II $N$-body simulation. In particular, we 
select both binary haloes in analogy to our MW-M31 system and haloes 
that are well isolated to see whether our model performs differently 
for objects in varying environments. This is relevant for the 
Local Group system which contains two massive galaxies. The much 
larger sample also enables us to investigate whether there are any 
physical variables that affect our model performance, such as the 
halo mass, shape, local environment and mass assembly histories. 

Real observations use stars as tracers. To study stellar tracers, we 
use an additional sample of haloes from a set of hydrodynamical 
simulations of the local group. In \cite{han2015a} and \cite{2015MNRAS.453..377W}, 
the star particles are selected from $N$-body simulations using the 
particle-tagging technique \citep{2010MNRAS.406..744C}. In this work, 
we instead use stars directly produced from the hydrodynamical 
simulations. This allows us to make a more realistic assessment of 
the dynamic status of stars, as well as an assessment of the 
reliability of particle-tagging technique for dynamical applications. 
Results from these hydrodynamical simulations are compared to 
their corresponding dark matter only runs to study the influence of 
baryonic physics in dynamical modelling. 

\cite{han2015a} and \cite{2015MNRAS.453..377W} have reported that both 
statistical and systematic errors tend to be aligned along a direction 
of anti-correlation between halo mass and concentration parameters even 
when the statistical errors are controlled to be much smaller 
than the systematic uncertainties. It has been argued that 
this is probably due to an underestimate of statistical errors. Particles 
sharing similar orbits in streams are highly correlated in phase space, 
and the true degree of freedom contributed by phase independent particles 
could be much smaller than the total number of particles. However, 
the small sample size used in the two previous studies prevents further 
investigations along this line. With our larger sample of haloes, we 
provide further support to this interpretation.

We introduce the set of simulations and tracers used for our analysis 
in Sec.~\ref{sec:mock}. The method of \cite{han2015a} is briefly 
summarised in Sec.~\ref{sec:method}. Results based on the large 
sample of haloes from the Millennium~II simulation and based on star 
particles in hydrodynamical simulations is presented in 
Sec.~\ref{sec:MRII} and Sec.~\ref{sec:LG}, respectively. In particular, 
in Sec.~\ref{sec:MRII} we investigate the relation between statistical 
and systematic errors. We investigate whether there are any hidden 
physical variables that systematically affect our model performance 
in Sec.~\ref{sec:depen}. We discuss the origin of parameter 
correlations in Sec.~\ref{sec:correlation}. A Navarro-Frenk-White 
profile \citep[][hereafter NFW]{1996ApJ...462..563N, 1997ApJ...490..493N} 
is often adopted to parametrize the halo density profile. Sec.~\ref{sec:nfw} 
is devoted to study how such a parametrization affects the results.

\section{Simulations and Tracers}
\label{sec:mock}
Our halo samples are selected from three different simulations as detailed 
below. Throughout this paper, we do not include particles belonging to 
subhaloes in our tracer sample. A thorough discussion of the 
influence of subhaloes can be found in \cite{han2015b}.

\subsection{Millennium~II}\label{sec:MRIIhalo}
In our analysis, we use a large sample of haloes selected from the 
Millennium~II Simulation \citep[][hereafter MRII]{2009MNRAS.398.1150B}. 
MRII is a dark matter only simulation with a box size of 100~$\mathrm{h^{-1}}$Mpc 
and a particle mass of $6.9\times10^6\mathrm{h^{-1}M_\odot}$. The cosmological 
parameters follow those from the first year WMAP result
\citep[][$\Omega_\mathrm{m}=0.25$, $\Omega_\Lambda=0.75$, $\mathrm{h}=0.73$, 
$n=1$ and $\sigma_8=0.9$]{2003ApJS..148..175S}. 

We select both isolated and binary haloes from MRII. Firstly, we 
identify a parent sample of haloes whose masses are analogous to Milky 
way, i.e., $0.5\times10^{12}<M_{200}<2.5\times10^{12}\mathrm{M_\odot}$ 
\footnote{We use $M_{200}$ to denote the mass of a spherical region with 
mean density equal to 200 times the critical density of the Universe.}. 
To select haloes that are well isolated, we require that all companions within 
a sphere of 2~Mpc are at least one order of magnitude smaller in 
$M_{200}$. For binary haloes, we make the selection in analogy to the MW 
and M31 system. The two haloes are required to be separated by a distance 
of 500 to 1000~kpc, and for a sphere centred on the mid-point of the two 
haloes, with a radius of 1.25~Mpc, all companions within the sphere 
should be less massive than the smaller of the two. In the end 
we have 658 isolated haloes and 336 binary haloes (or 168 pairs). 

Inside each halo, we use dark matter particles that do not belong to any 
bound subhaloes and are within the virial radius\footnote{The radius 
within which the mean density is 200 times the critical density of the 
universe.}, $R_{200}$, as tracers. The mean number of such dark matter 
particles in each halo is about $10^5$.

\subsection{The \textsc{apostle} simulations}
\label{sec:LGhalo}
\textsc{apostle} stands for A Project of Simulations of The Local Environment 
\citep{2015arXiv150703643F,2015arXiv151101098S}. It consists of a 
suite of 12 high resolution cosmological Smoothed-particle hydrodynamics
(SPH) simulations of Local Group-like environments selected from 
large cosmological volumes of a $\Lambda$CDM universe. They are then 
re-simulated with three different levels of resolution using the 
\textsc{eagle} \citep{2015MNRAS.450.1937C,2015MNRAS.446..521S} hydrodynamics 
code. The high resolution region of each simulation contains a pair of 
galactic haloes corresponding to the MW and M31. The underlying cosmology 
of \textsc{apostle} is that of WMAP7 \citep[][$\Omega_\mathrm{m}=0.272$, 
$\Omega_\Lambda=0.728$, $h=0.704$, $n=0.967$ and $\sigma_8=0.81$]{Komatsu2011}. 
The particle mass of the lowest resolution run is comparable to the 
intermediate resolution \textsc{eagle} run. The intermediate and high level
runs have mass resolutions higher by factors of 12 and 144 respectively, 
but the high resolution runs are not yet completed for all twelve pairs.

For our analysis we choose to use the suite of intermediate resolution 
simulations. Each halo in the intermediate level run contains about 
$\sim 10^4$ to $\sim 10^5$ star particles in the stellar halo that are 
not bound to any satellites or subhaloes, and these star particles are 
used as our dynamical tracers. The mass of dark matter particle in these 
simulations range between $\sim 3\times10^5$ to $\sim 4\times 10^5
\mathrm{M_\odot}$.  We label the 12 simulations as V1 to V6 and S1 to S6.
\footnote{The 12 simulations are called AP-1 to AP-12 in the same order as V1 
to V6 and then S1 to S6 in the introductory \textsc{apostle} paper \citep{2015arXiv150703643F}.}
For the intermediate resolution simulations, each of the six ``V'' simulations 
contains two haloes that are in two separate Friends-of-Friends groups. The 
two haloes in each of the ``S'' simulations are all in the same 
Friends-of-Friends group and linked together due to particle bridges.

We also use dark matter particles in these haloes as tracers. Each 
pair of \textsc{apostle} haloes is also simulated in a corresponding dark 
matter only (hereafter DMO) run. We refer to haloes in the hydrodynamical 
run as \textsc{apostle} haloes, while the DMO versions are referred to as 
DMO haloes or runs. For DM tracers, we analyse both the \textsc{apostle} 
and DMO haloes side by side. Given the higher resolution of 
\textsc{apostle} haloes, we choose to use subsets of dark matter 
particles, which are comparable in sample size to MRII haloes. We 
have explicitly checked that our results are not affected by randomly 
selecting different subsets. 
 
\subsection{The Aquarius and the mock stellar halo simulations}
\label{sec:Aqhalo}
The Aquarius simulations are N-body simulations in a standard 
$\mathrm{\Lambda}$CDM cosmology \citep{Springel_2008}. Cosmological 
parameters are those from the first year data of WMAP 
\citep{2003ApJS..148..175S}. Our work uses the second highest 
resolution level of the Aquarius suite, which corresponds to a 
particle mass of $\sim10^4\mathrm{M_\odot}$. The simulation 
includes six dark matter haloes with virial masses spanning 
$0.87\times10^{12}$ to $1.8\times10^{12}\mathrm{M_\odot}$. 
Aquarius haloes are locally isolated with no nearby massive 
companions within 1~Mpc. 

The mock stellar halo catalogues are constructed based on the 
particle tagging method developed by \cite{2010MNRAS.406..744C}, 
to which we refer the reader for further details. In brief, 
the method associates the predicted star formation from the 
Durham Galaxy formation models \citep[\textsc{GALFORM};][]
{2000MNRAS.319..168C,2008ApJ...673..215F} with the most 
bound portion of the host dark matter subhalo. The spatial 
distribution and velocities  of newly formed stars
are initially represented by the 1\% most bound 
dark matter particles within the host subhalo. 
Later these particles can be stripped to form
a stellar halo. This approach is 
based on the knowledge that stars are much more dynamically 
bound and radially concentrated than dark matter. The method 
can reproduce well the size-luminosity relations of MW 
satellite galaxies. 

Previously, \cite{han2015b} have analysed five out of 
the six haloes (labelled halo A to halo E in the Aquarius convention) 
with the same \textsc{oPDF} method, using both the tagged star 
particles and dark matter particles as dynamical tracers. In our 
analysis, we do not directly use the Aquarius suite of simulations, 
but we make direct comparisons with \cite{han2015b}. There are 
2 to $5\times10^5$ star particles in the Aquarius stellar haloes.
The mock stellar halo catalogues developed by \cite{2010MNRAS.406..744C} 
have been widely used before by \cite[e.g.][]{2011ApJ...733L...7H,
2011MNRAS.417.2206C,2013MNRAS.436.3602G,2015MNRAS.446.2274L,
2015MNRAS.453.2830M,2015AJ....150..160B,2015arXiv150206371L} for 
the study of Milky Way like stellar haloes and the comparison to 
real observations.

\section{Methodology}
\label{sec:method}
The method we are going to investigate is the orbital probability 
density function (hereafter \textsc{oPDF}) method developed by 
\cite{han2015a}. This is a general method in the sense that it 
requires minimal assumptions about the distribution function of 
the system. In fact the only assumption about the tracer 
distribution is its time independence, a fundamental assumption 
underlying any steady-state model of the system. As a result, the 
systematic uncertainty revealed by the oPDF method can be 
readily interpreted as deviations from a steady state. It also 
represents the minimum systematic uncertainty expected from any 
steady-state dynamical model as they typically invoke additional 
assumptions about the functional form of the distribution function. 

The starting point of the method is that in a steady-state system, 
the probability of observing a particle at a given position is 
proportional to the time it spends at that position. If we label 
the position of each particle by its travel time from a reference 
point, $t(\vec{r})$, we can define a phase angle (also known 
as the radial action angle) as
\begin{align}
\theta(\vec{r})&=\frac{t(\vec{r})}{T_r},
\end{align}
where $T_r$ is the period of the orbit. The steady state requirement immediately implies 
\begin{equation}
\ud P(\theta|\rm{orbit})=\ud \theta.\label{eq:theta_distr}
\end{equation} That is, the particles are uniformly distributed in phase angle along 
each orbit. Equation~\ref{eq:theta_distr} can be derived from the time independent collisionless Boltzmann 
equation (\citealp{han2015a}, see also Appendix~\ref{sec:equiv_continuity} for an alternative derivation), 
and holds for each of the three action-angles.
Particles following this distribution are referred to as fully \emph{phase mixed}. 

In a spherical potential, the orbital distribution can be alternatively expressed via a 
radial coordinate as
\begin{equation}
\ud P(r|E,L)=\frac{1}{T_r}\frac{\ud r}{|v_\mathrm{r}|}, 
\label{eqn:radialprof}
\end{equation}
where $P(r|E,L)$ is the probability of finding a tracer object at radius, 
$r$, given its binding energy, $E$, and angular momentum, $L$. The radial velocity at 
any radius, $r$, can be predicted as 
\begin{equation}
 v_\mathrm{r}=\sqrt{2\Phi(r)-2E-L^2/r^2}. 
\end{equation}
Equation~\ref{eqn:radialprof} is equivalent to the Jeans 
Theorem. In principle, our method can be generalized to higher dimension 
and an arbitrary potential, as briefly discussed in section~6 of \citep{han2015a}.

Starting from the observed position $\vec{r}_i$ and velocity $\vec{v}_i$ of a 
tracer particle $i$, one can obtain its orbital parameters, 
$E_i=-\left(\vec{v}_i^2/2+\Phi(r_i)\right)$ and $L_i=|\vec{r}_i\times \vec{v}_i|$, 
for any assumed potential $\Phi(r)$. Combining the contributions from all the 
particles, the overall radial distribution of the tracers can be predicted as
\begin{equation}
 P(r)=\frac{1}{N}\Sigma_i P(r|E_i,L_i).
 \label{eqn:radialprof2}
\end{equation}
Requiring that the predicted radial profile matches the observed profile, we can solve 
for the true potential of the system. In practice, the solution is found in a 
statistical manner. If we bin the data radially into $m$
bins, the expected number of particles in the $j$-th bin is given by
\begin{equation}
\hat{n}_j= N\int_{r_{\mathrm{l},j}}^{r_{\mathrm{u},j}} \frac{\mathrm{d} P(r)}{\mathrm{d} r} \mathrm{d} r,
\end{equation}
where $r_{\mathrm{l},j}$ and $r_{\mathrm{u},j}$ are the lower and
upper bin edges. The binned radial likelihood is given by:
\begin{align}
 {\cal L}&=\prod_{j=1}^{m} \hat{n}_j^{n_j} \exp(-\hat{n}_j)\\
 &=\exp(-N) \prod_{j=1}^{m} \hat{n}_j^{n_j},
\end{align}
where $n_j$ is the observed number of particles in the $j$-th bin. The best-fitting 
potential is defined to be the one that maximises this likelihood. The $1\sigma$ 
confidence region of the model parameters can also be obtained by scanning for the 
likelihood contour with $\Delta \ln {\cal L}=1.15$ (for two model parameters) from 
the likelihood peak.

Practical applications of the above method require a parametrization of the potential 
profile, $\Phi(r)$. To segregate the effect of a poor parametrization from other 
systematics, our default in the following analysis is to parametrize $\Phi(r)$ using 
a template generalised from the true profile. Explicitly, the model profile is 
parametrized as $\Phi(r)=A \Phi_{\rm true}(B r)$, where $\Phi_{\rm true}(r)$ is the 
true profile extracted from the simulation, and $A$ and $B$ are free parameters 
determining the normalisation and radial scale. In addition to the template 
parametrization, we also present results using the popular and practical  
parametrization of an NFW profile. Comparisons between the two reveal how 
much the assumed parametrization affects the modelling.

The parameters $A$ and $B$ can be equivalently converted to the mass and 
concentration parameters of the halo following \citet{han2015b}. In the following 
analysis, we always work in the mass and concentration parameter space and present 
the results accordingly. The mass, $M_{200}$, is defined to be the total mass 
inside the virial radius, $R_{200}$, the radius enclosing an average density of 
200 times the critical density of the universe. The concentration, $c_{200}$, is 
defined as $R_{200}/r_\mathrm{s}$, where $r_\mathrm{s}$ is the radius at which 
the density profile has a logarithmic slope of $-2$.

\section{Results from MRII: the irreducible uncertainty of steady-state models}
\label{sec:MRII}

We at first analyse all binary and isolated haloes from MRII as a whole, 
without distinguishing between them. This maximises the sample size and helps 
us to robustly quantify the scatter of the systematic errors. Throughout this 
section, we use dark matter particles within $R_{200}$ but outside $20$~kpc 
from the host centre as tracers. True potential templates are used in the modelling.

\begin{figure*}
\epsfig{figure=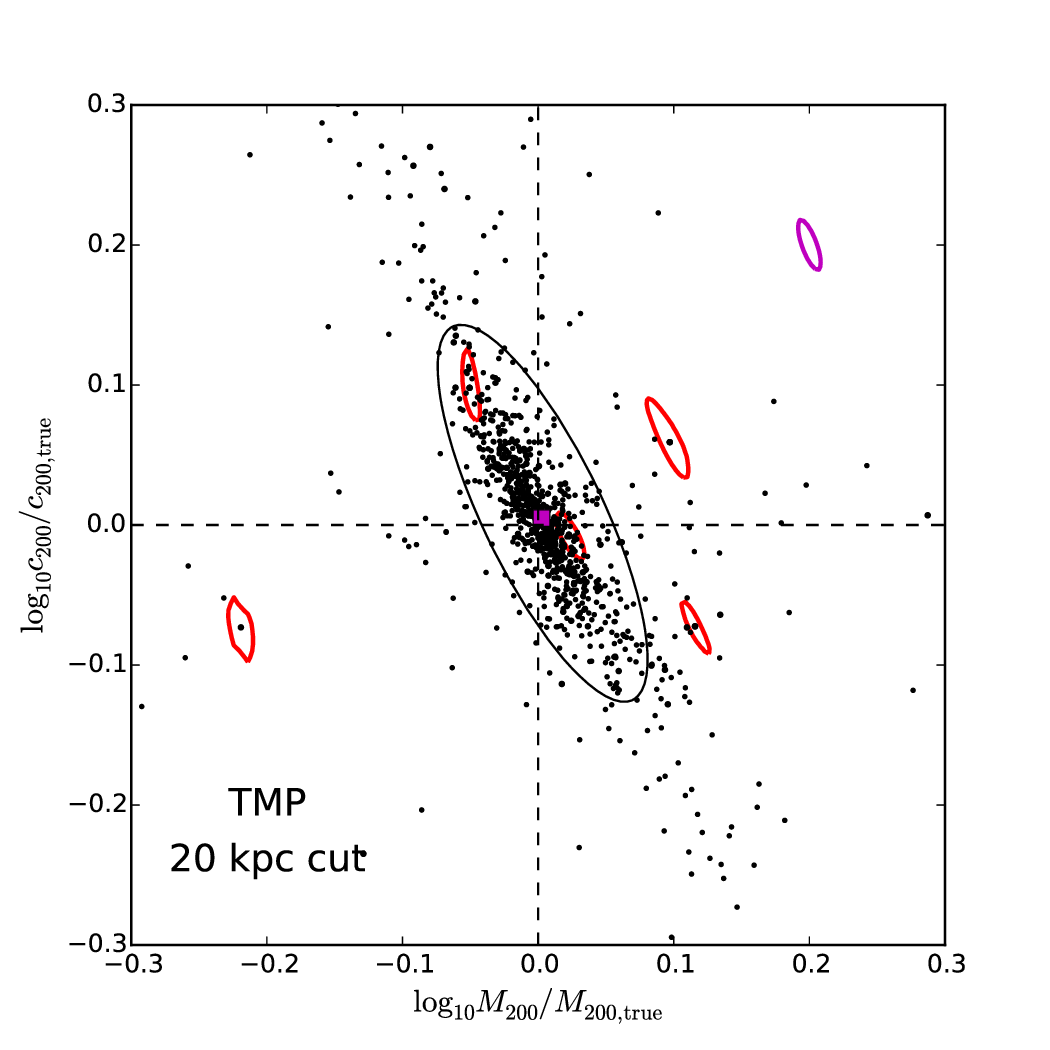,width=0.49\textwidth}
\epsfig{figure=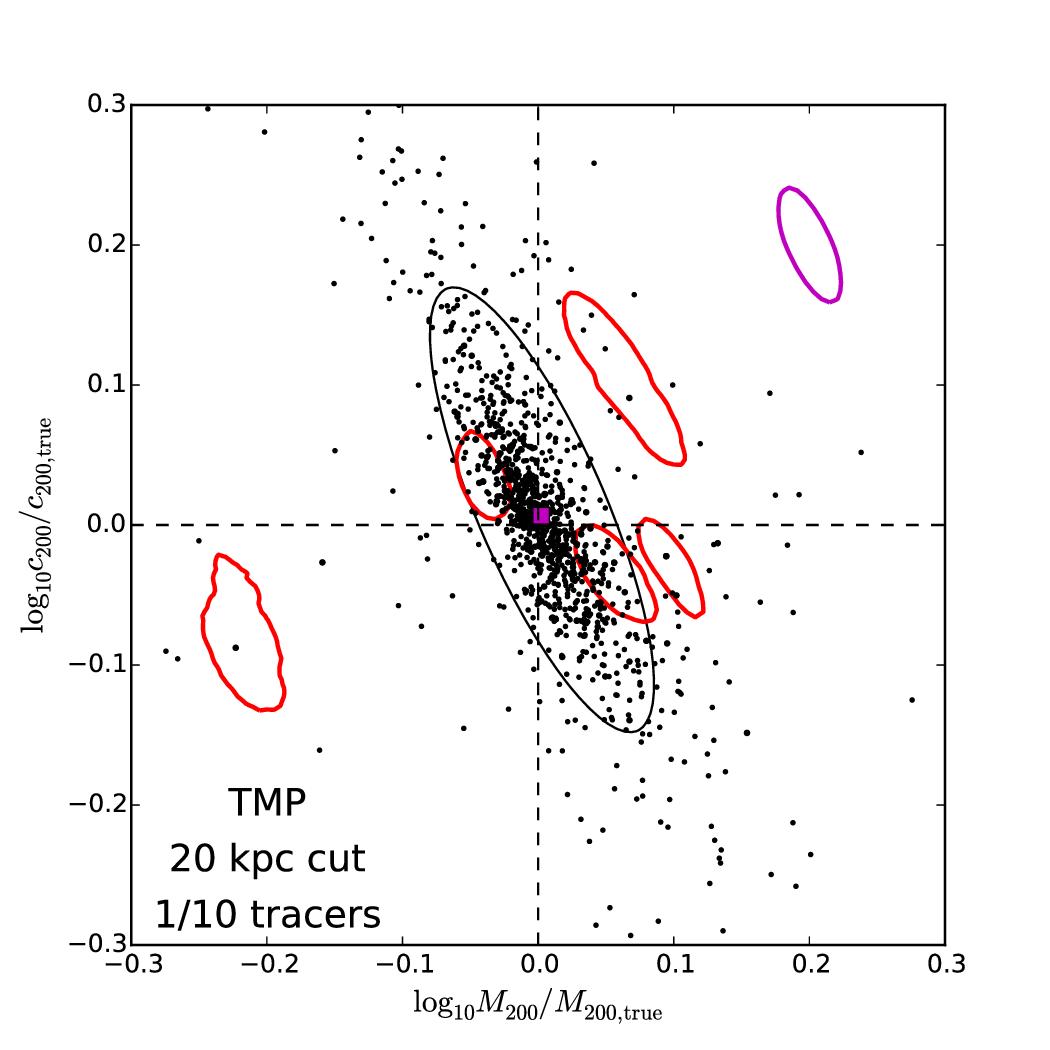,width=0.49\textwidth}%
\caption{{\bf Left:} Best fit halo mass ($x$-axis) and concentration 
($y$-axis) in units of their true values, for haloes in the MRII sample using 
dark matter particles as tracers. We adopt an inner radius cut of 20~kpc and 
model the underlying potential with templates (TMP). Each dot represents the 
fit to one halo. Horizontal and vertical black dashed lines mark the equality 
between best fit and true parameters. The magenta solid square is the average 
parameter of all the haloes and is very close to the origin. The black 
ellipse marks the $1~\sigma$ scatter of all the measurements. Red contours 
show the 1~$\sigma$ statistical errors given by likelihood contours for five 
randomly selected haloes. The magenta ellipse in the top right corner shows 
the average 1~$\sigma$ contour over the entire sample of haloes. {\bf Right:} 
Same as the left, but each halo is down-sampled by a factor of 10 in the number 
of tracer particles. The statistical errors are shown for the same five haloes. 
There is a significant increase in statistical errors, whereas the scatter in 
systematic errors only becomes slightly larger.} 
\label{fig:pairiso20} 
\end{figure*}

The left panel of Fig.~\ref{fig:pairiso20} shows the fitted parameters of haloes 
in the MRII sample. On average, the fits are unbiased. However, each individual 
fit could deviate significantly and stochastically from the true parameters. To 
quantify the scatter of these deviations, we estimate the covariance of the 
points and plot the 1~$\sigma$ confidence region with the black ellipse, assuming 
a Gaussian distribution of the points with the estimated covariance.\footnote{ 
Although we have chosen the axis range to be from -0.3 to 0.3, there are a small 
fraction of measurements outside this range. We at first calculated the ellipse 
size and orientation based on the covariance matrix of all converged measurements. 
We then excluded the most biased measurements using 3-$\sigma$ clipping and 
plotted in Fig.~\ref{fig:pairiso} the ellipse based on the remaining measurements.}
It indicates a scatter of about 25\% in $M_{200}$ and 40\% in $c_{200}$, which 
is much larger than the typical statistical error of each individual fits (red 
contours). Most importantly, the covariance of the points appear to be a 
scaled version of the statistical noise. They both align in a direction of 
anti-correlation between $M_{200}$ and $c_{200}$. This is significant 
because in principle systematic errors can happen along any direction in 
parameter space, regardless of the direction of the statistical errors. 

A viable explanation is that the statistical noise is underestimated, as 
already proposed in \cite{2015MNRAS.453..377W} and \cite{han2015b}. This is 
expected due to the preponderance of correlated phase-space structures such 
as streams and caustics in the simulated haloes~\citep{Helmi&White99, 
Vogelsberger&White11}. Particles inside each stream share similar orbits, 
but are highly correlated in their orbital phases~\citep{han2015b}. 
As a result, the number of independent particles, or the effective sample 
size determining the statistical noise, is smaller than the actual 
sample size, leading to an increase in the statistical noise compared to a 
fully independent sample of particles. However, as long as the streams are 
uncorrelated with each other, they are not expected to bias the fit 
statistically. This is consistent with Fig.~\ref{fig:pairiso20} that 
shows the average parameter values of all the haloes are very close to 
being unbiased.

Note that the existence of correlated phase-space structures also means deviations 
from a steady state, because the phase-space density evolves as the structures 
move. So the underestimated statistical noise is indeed a source of systematic 
uncertainty for the steady-state assumption. More importantly, such 
an uncertainty cannot be reduced by simply increasing the sample size of 
the tracer, because its size is determined by the effective number of phase 
independent particles intrinsic to each halo. This is shown in the right panel 
of Fig.~\ref{fig:pairiso20}, where we repeat our analysis after down-sampling 
the tracers of each halo by a factor of 10. It is clear that with the reduction 
in the tracer population size, the statistical errors become significantly 
larger, reflecting a scaling with $1/\sqrt{N}$ where $N$ is the sample size.
\footnote{This estimate does \emph{not} rely on the assumption of Poisson errors 
but is a leading order result following the standard likelihood error analysis. 
For $N$ equal size samples following identical independent distributions, the 
combined log-likelihood function is simply $N\ln L_0$ where $L_0$ is the likelihood 
of a single sample. As a result, the Hessian matrix of the log-likelihood 
increases by a factor of $N$, so that the combined error reduces by a factor 
of $1/\sqrt{N}$ compared with that of a single sample.} On the other hand, 
the scatter of these measurements only increases slightly and obviously 
does not follow the $1/\sqrt{N}$ scaling. Note the measured scatter includes 
contributions from both the systematic and the statistical uncertainty, 
so that the slight increase is expected even though the systematic uncertainty
remains largely unchanged. This has important implications for real 
observations. It means the measurement uncertainty from steady-state 
models saturates to an \emph{irreducible} intrinsic uncertainty once the sample 
size becomes much larger than the effective number of independent particles.

Assuming the black ellipse and the magenta 1-$\sigma$ error contour 
would have the same size if the true effective number of phase independent 
particles are properly considered, we can make a rough estimate of the 
effective numbers of phase-independent particles \footnote{A possible 
dependence on the spherical assumption has been ignored here.}. 
In the left plot, the size of the black ellipse is roughly 10 times larger 
than the size of the magenta contour, which means the effective number of 
particles can be roughly estimated as $1/10^2$ times the total number of 
particles. The mean number of tracer dark matter particles in these MRII 
haloes is about $10^5$. Thus the mean effective particle number is 
$N_{\rm eff}\sim 1000$. Applying the same analysis to the right panel 
leads to approximately the same result. This again confirms our interpretation 
that the irreducible bias is controlled by the intrinsic number of independent 
particles in the halo. We return to discussions of the effective 
particle number in Sec.~\ref{sec:merger}. Also note we have so far 
ignored other possible sources of systematic uncertainty in our modelling and the effective 
number of particles are estimated by assuming the uncertainties 
are dominated by violations of the steady state assumption. More detailed 
discussion will be made in Sec.~\ref{sec:depen}, including investigating 
the dependence on deviations from spherical symmetry (Sec.~\ref{sec:shape}).

\section{Stellar tracers in \textsc{apostle}}
\label{sec:LG}

The large sample of MRII simulation has enabled us to quantify and 
understand the intrinsic scatter in systematic errors. However, MRII 
does not have baryons and only dark matter particles can be used as 
tracers. In this section we now apply \textsc{oPDF} to star 
particles in \textsc{apostle}. This enables us to more closely connect to 
 real observations, where only luminous objects can be used for 
dynamical modelling. Still we will focus on using the true potential 
templates.

\subsection{Halo stars in \textsc{apostle} as tracers}
\label{sec:star}
\begin{figure} 
\epsfig{figure=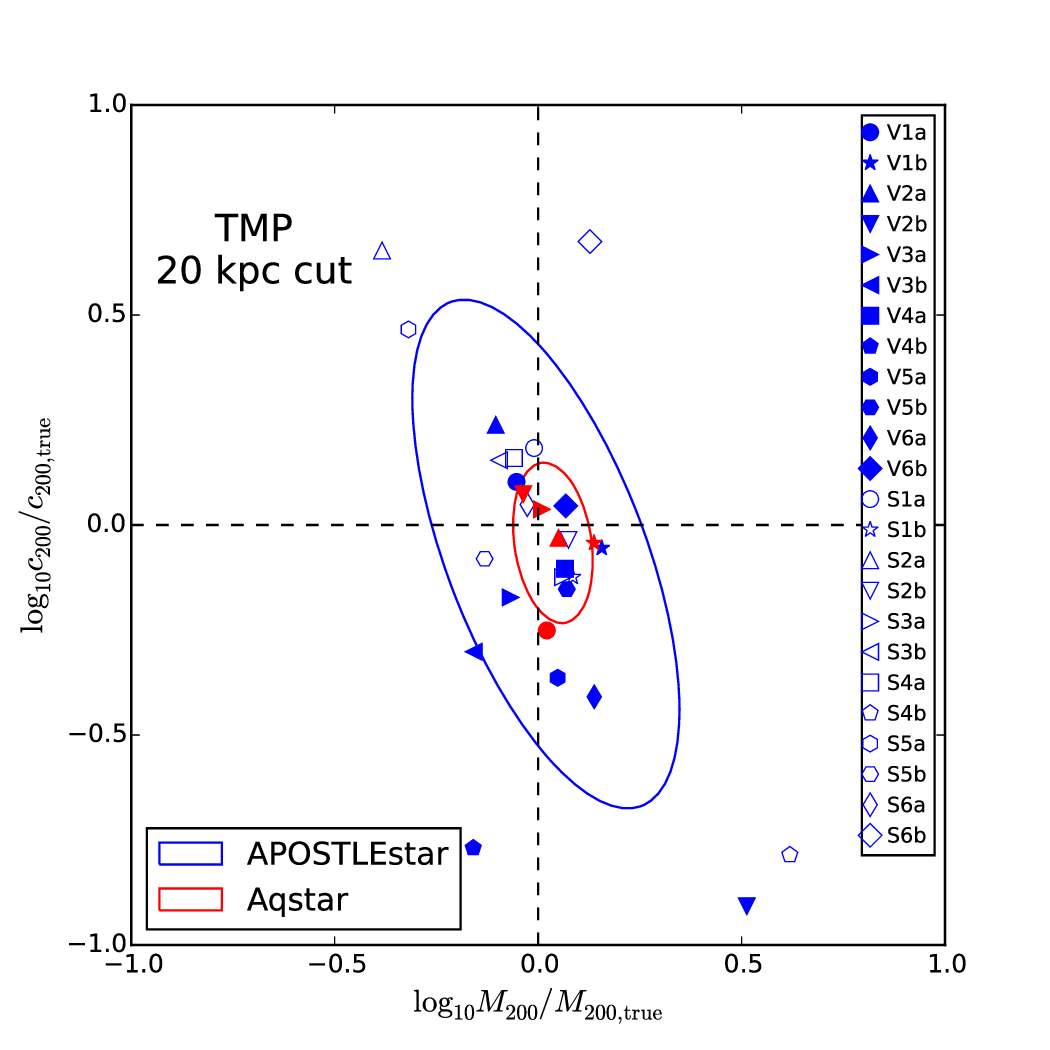,width=0.49\textwidth}%
\caption{The best fit $M_{200}$ and $c_{200}$ in units of their true values using 
star particles in \textsc{apostle} haloes as tracers (blue). The meaning of symbols are 
indicated by the legend. For example, V1a and V1b refer to the M31 and MW analogies 
in the V1 simulation volume. Red symbols are results from Aquarius haloes. Blue and 
red ellipses mark the 1~$\sigma$ estimated from the covariance of the measurements.} 
\label{fig:star} 
\end{figure}

\begin{figure} 
\epsfig{figure=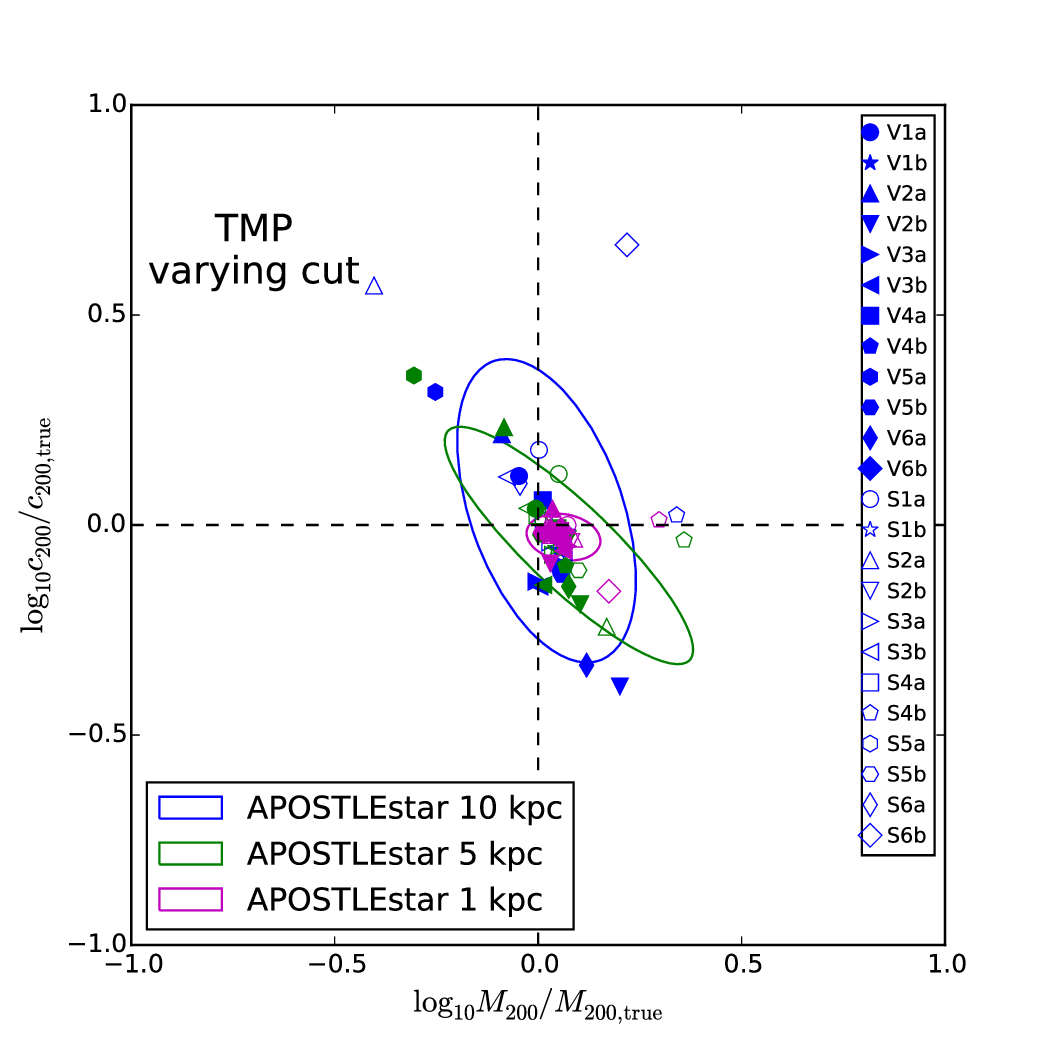,width=0.49\textwidth}%
\caption{As Fig.~\ref{fig:star}, but different inner radius cuts 
(10, 5 and 1~kpc) are adopted to investigate the effect of stars 
in the very central region that are mainly formed in situ in the 
disc. The results are shown in different colours as labelled.} 
\label{fig:insitu} 
\end{figure}

We again at first use star particles located between 20~kpc and $R_{200}$ and 
not bound to any subhaloes as tracers. This is to avoid the central 
disc component, which is usually believed to violate the spherical 
assumption and for real data stellar tracers within 20 or 15~kpc are 
often not used \citep[e.g.][]{2005MNRAS.364..433B,2008ApJ...684.1143X,
2010ApJ...720L.108G,2012MNRAS.424L..44D}. We postpone more detailed 
discussion regarding the inner tracers to Sec.~\ref{sec:insitu}. 
Fig.~\ref{fig:star} shows the best fit $M_{200}$ and $c_{200}$ 
versus their true parameter values in \textsc{apostle} (blue symbols). 
We can see a very large scatter in the best fit parameters.
 
We overplot as red symbols the best fit parameters based on mock halo 
stars in the five Aquarius haloes.\footnote{The original analysis in \cite{han2015b} 
adopted an inner radius cut of 10~kpc. Here we have repeated the calculation for 
Aquarius haloes adopting an inner radius cut of 20~kpc, to be consistent with 
the cut adopted for \textsc{apostle} haloes. This only makes the systematic uncertainty
slightly larger than that of \cite{han2015b}. There is a 20\% to 50\% systematic 
uncertainty} in the best fit parameters of the Aquarius haloes. This appears to be 
significantly smaller than the systematic uncertainties for the \textsc{apostle} 
haloes, which can be as large as a factor of three in $c_{200}$ and a factor of two in $M_{200}$.

Since baryons in \textsc{apostle} were directly simulated whereas stars in Aquarius 
are inserted by tagged using the most bound subset of dark matter particles, 
there might be some differences between the two approaches that complicate direct 
comparison. However, \cite{2015arXiv150206371L} have compared the properties of 
stars in the stellar halo between hydrodynamical simulations and stars created 
by particle tagging. They find that if particles are regularly tagged throughout 
the evolution of the  galaxy, tagging can reproduce well the density profiles, 
binding energy and angular momentum distributions of halo stars in hydrodynamical 
simulations at low redshifts. Despite this some uncertainty remains as to whether 
the smaller scatter of Aquarius haloes is mainly due to differences between the 
particle tagging approach and hydrodynamical simulations. The way to verify this 
would be to construct mock stellar halo catalogues for \textsc{apostle} haloes 
and make direct comparisons. Such mock catalogues are not yet available for 
\textsc{apostle} haloes. Nevertheless, we note that as there are only five 
Aquarius haloes used for our analysis, it is quite possible that these haloes 
happen to be good cases simply due to statistical fluctuations. 

To quantify the statistical significance of the difference, we perform
a one dimensional Kolmogorov-Smirnov (K-S) test to estimate the probability 
that the fitted parameters of the two samples are drawn from the same distribution.
This is done separately for the two parameters. The resulting p-values are 0.3656 
for $M_{200}$ and 0.6196 for $c_{200}$, This means that we have 36.56\% (61.96\%) 
probability of observing an equal or even larger difference in the two mass 
(concentration) samples, which implies that the two samples are not statistically 
distinguishable for either parameter. Repeating the one dimensional K-S test for 
the parameter combination corresponding to anti-correlation direction we find a 
p-value of 0.6506, again compatible with the smaller scatter of Aquarius haloes 
appearing by chance due to the small sample size.

With the larger sample of \textsc{apostle} haloes, we found a significantly 
larger scatter in the best fit parameters compared to those based on Aquarius 
haloes. The nearly a factor of two scatter in $M_{200}$ and a factor of 
three scatter in $c_{200}$ is worrying for dynamical modelling of MW halo 
mass. As discussed above, this amount of systematic uncertainty is irreducible 
and applies to any steady-state models.

The typical size of the statistical errors is smaller than the symbol size 
in Fig.~\ref{fig:star}, and is about a factor of 0.03 of the scatter of 
the systematic uncertainty (blue ellipse). With an average sample size 
of $4.5\times10^4$ star particles in each \textsc{apostle} halo, we find the 
effective number of independent star particles (see Sec.~\ref{sec:MRII}) to 
be $N_{\rm eff}\sim 0.03^2\times 4.5\times10^4 \approx 40$. Recall that in Sec.~\ref{sec:MRII} 
we estimated the effective number of dark matter particles in MW-like haloes 
to be about 1000. This means, star particles in our hydrodynamical simulations 
are far more correlated than dark matter particles. This is not surprising, 
because we know stars have much higher binding energy than dark matter and
are more centrally concentrated. They are less relaxed and less phase 
mixed than dark matter. In addition, we expect about 30\% to 40\% of 
dark matter particles are smoothly accreted instead of being accreted 
as part of a bound substructure \citep[see details in][where a detailed 
investigation on possible dependencies on the resolution has been made]{2011MNRAS.413.1373W}, 
whereas we have checked that more than 80\% of star particles beyond 
20~kpc of the halo centre are stars stripped from satellites 
in the simulation. Smoothly accreted particles could be much less 
correlated in phase space than particles stripped from subhaloes, 
which are expected to cluster in phase space around each subhalo. We 
return to the discussion of the implications on the effective particle 
number and the connection to halo merger histories in 
Sec.~\ref{sec:merger}. Note again we have not tested other sources 
of bias including the assumption of spherical symmetry, and we defer this 
to Sec.~\ref{sec:depen}.

\subsection{Stars formed in situ}
\label{sec:insitu}

\begin{figure} 
\epsfig{figure=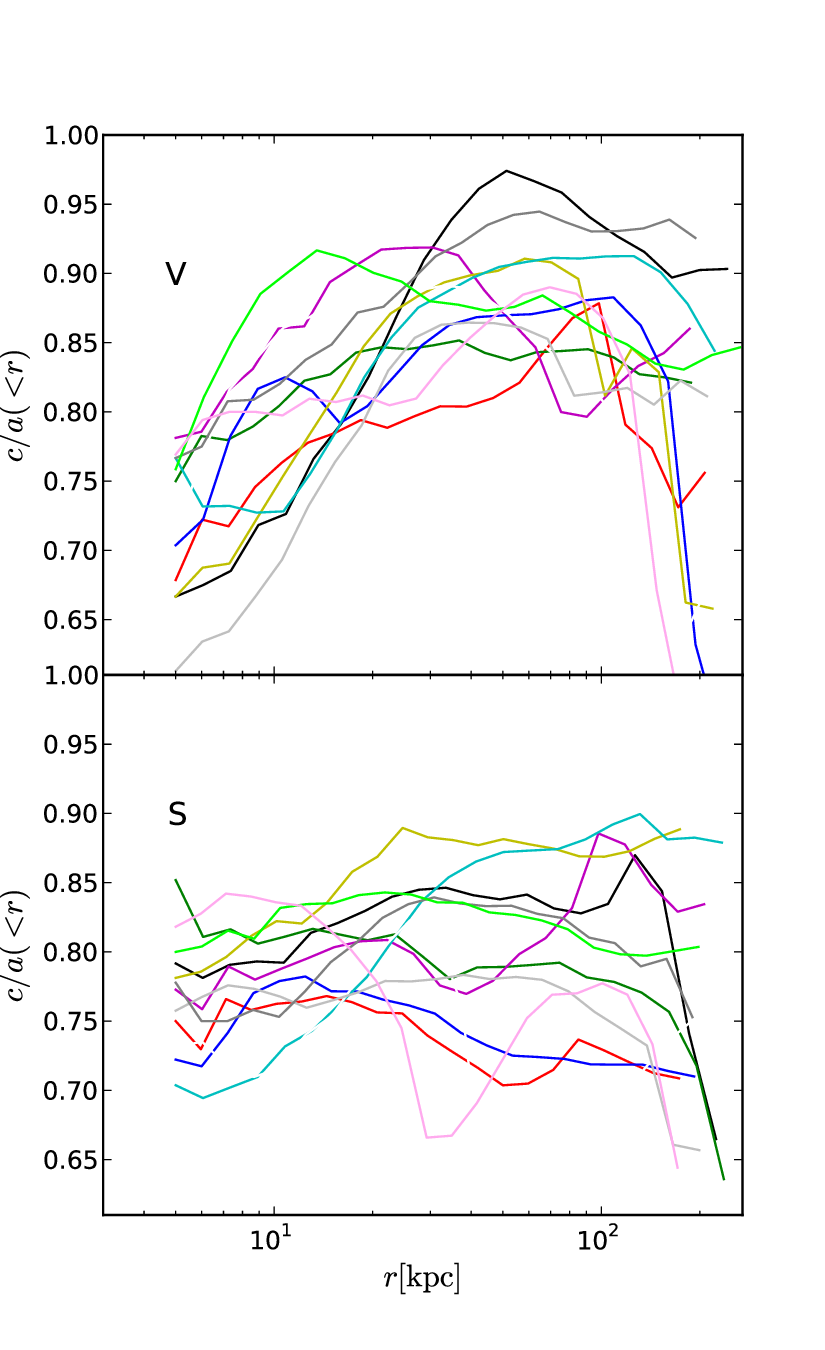,width=0.49\textwidth}%
\caption{The minor to major axis ratio, $c/a$, of the mass weighted 
inertial tensor within radius, $r$, obtained by considering all particles 
(stars, dark matter, gas and black hole).} 
\label{fig:caradius} 
\end{figure}

In Fig.~\ref{fig:star}, we have excluded star particles within 20~kpc of 
the halo centre. We now try to see how the result would change if we 
include stars in the very central region, which includes the disc component 
and is mostly formed in situ. In Fig.~\ref{fig:insitu} we tried three 
different inner radius cuts, 1, 5 and 10~kpc. Surprisingly, the scatter 
becomes smaller with the decreasing inner radius cut. The 1~kpc inner 
radius cut gives very small scatter in the best fit parameters. This is 
inconsistent with the naive expectation that stars formed in situ in the 
central disc violate the spherical assumption and may introduce stronger 
systematic uncertainties.

Fig.~\ref{fig:caradius} shows the minor to major axis ratio, $c/a$, of the 
inertia tensor obtained from the mass distributions within different radii. 
It seems the haloes are still close to being spherical at 20~kpc, with $c/a$ 
mostly above 0.8. Within 20~kpc, $c/a$ decreases for most of the V haloes in 
the upper panel, but the values are mostly above 0.7. For S haloes in the 
lower panel, the values of $c/a$ are slightly smaller than for V haloes 
outside 20~kpc and are close to being flat within 20~kpc. This is possibly 
related to the fact that V haloes are in separate friends-of-friends groups, 
while S haloes are in the same group. This suggests the inner potential 
profiles are not extremely oblate or elongated in our simulation, and thus 
it helps to explain why the inclusion of stars in the very inner 
region does not make the fits worse. However, it remains to be seen whether 
the spherical potential in \textsc{apostle} is a realistic representation of 
the real Galaxy, or is instead a result of the implemented subgrid physics 
which may not be realistic enough to model the baryon distribution in the 
very inner region, though this is currently the best we can achieve. The 
simulated potential may not reflect the true potential profile in the central 
region of the MW, where a vertical X-shaped structure has been detected 
\citep[e.g.][]{2012ApJ...757L...7L}. More details about the shape and 
alignment of MW-like galaxies, haloes and their satellite systems in 
\textsc{eagle} can be found in \cite{2015MNRAS.453..721V,2015MNRAS.454.3328V} and 
\cite{2016arXiv160501728S}.

Nevertheless, the above results lead us to the interesting conclusion that 
stars in the very central region must be very relaxed, in order to achieve 
the very small scatter in the systematic errors. This has 
previously been  discussed in \cite{2005ApJ...635..931B}. Combined with proper 
modelling of the underlying potential profiles, these inner stars can help 
to better constrain halo properties and mass profiles. This is not surprising, 
as we know stars at smaller radii are expected to have higher binding energy, 
while \citet{han2015b} has found that particles at smaller radii and with 
higher binding energies are more relaxed. This is also consistent with the 
shorter dynamical time in the centre.  We will have more discussion 
in Sec.~\ref{sec:nfw} to see what will happen if the underlying potentials 
are not properly modelled.

The effective numbers of phase-uncorrelated particles are about 90, 103 and 
730 for the 10, 5 and 1~kpc cuts. Apparently with the inclusion of particles in 
the very inner region, the number of phase-uncorrelated particles is significantly 
increased, which further supports the view that in-situ stars in the very central 
region are well phase mixed and much more relaxed.

\section{The causes of the systematic uncertainty}
\label{sec:depen}

In this section we look for hidden variables which may be responsible for the 
systematic uncertainties in the fits. Such an investigation could improve our understanding 
of the dynamical state of the haloes. If found, these variables could serve to 
predict the intrinsic uncertainty in the fit to each halo. 

To this end, we have investigated the following list of properties:
\begin{itemize}
\item halo environment, focusing on whether the halo is isolated or in a binary system;
\item halo shape, as quantified by the axis ratio;
\item The radial range of tracer particles;
\item halo merger history, quantified by the number of resolved progenitors/subhaloes;
\item pair separation and mass ratio for binary haloes;
\item halo mass;
\item halo concentration.
\end{itemize} 

We fail to discover obvious dependencies on halo mass, 
concentration and binary separations. It is possible that the 
dependencies are too weak to show up for MW-like haloes sharing a 
limited range in the halo mass and binary separation. However, we 
do see some dependence on environment, shape, and merger history 
of the halo, as expanded below.

For results in this section, we mainly focus on the large sample of 
MRII haloes due to its statistical power. In contrast to Sec.~\ref{sec:MRII}, 
an inner radius cut of 1~kpc is used to select dark matter tracers in MRII 
since we have seen that this inner cut leads to improved fits when star particles 
are used as tracers. A comparison between Fig.~\ref{fig:pairiso20} and the results 
below enables us to see whether the radial cuts of tracers systematically affect 
the systematic uncertainties in the fits. However, we have checked that the dependence 
on halo environment, shape, and merger history are not affected by the choice of 
inner radius cut.

\subsection{Binary versus isolated haloes}
\label{sec:pairiso}
\begin{figure}
\epsfig{figure=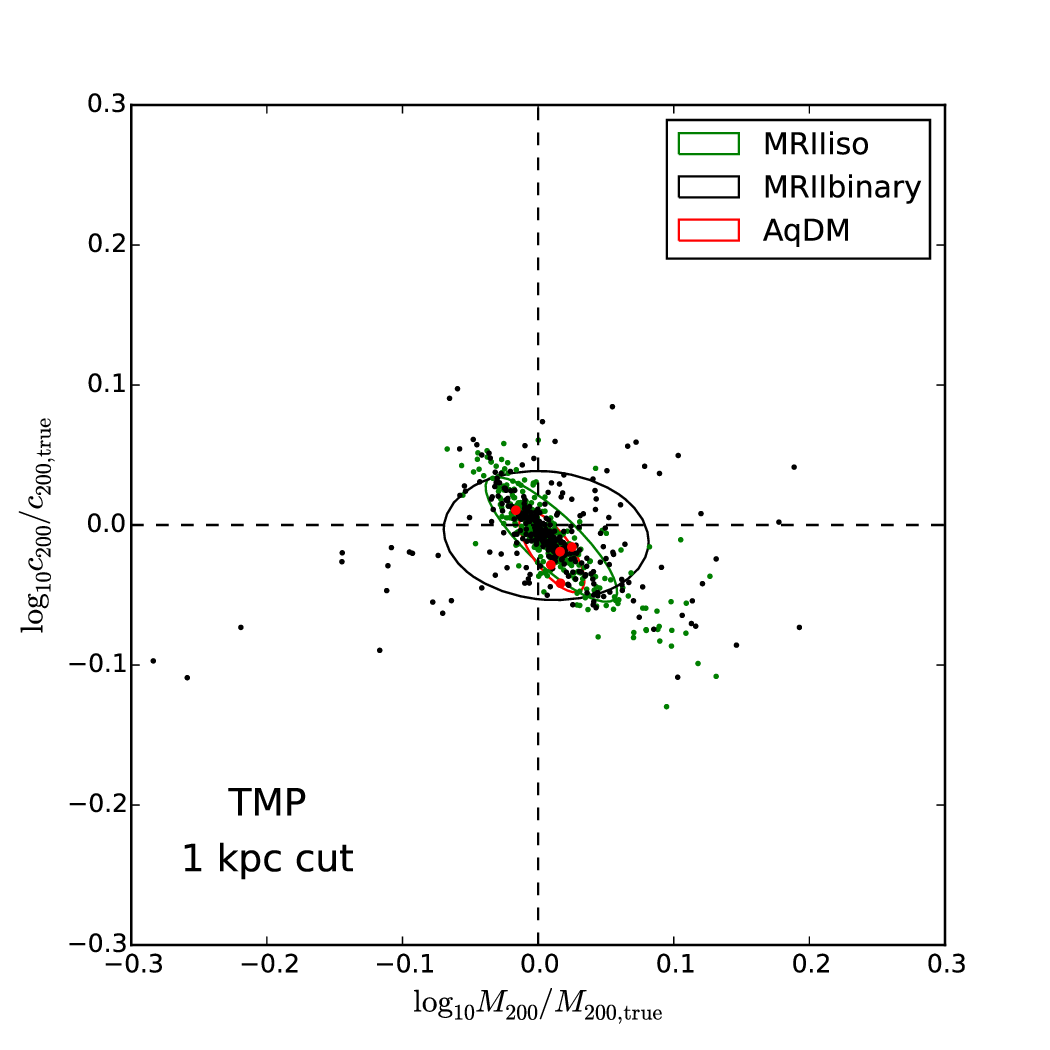,width=0.49\textwidth}%
\caption{The fits to binary (black) and isolated (green) galactic haloes in the MRII 
sample. Points are fits to individual haloes and the ellipses mark the $1~\sigma$ 
scatters. For comparison, red dots show fits to DM tracers in the Aquarius haloes. 
An inner radius cut of 1~kpc is adopted for all haloes.} 
\label{fig:pairiso} 
\end{figure}

\begin{figure*} 
\epsfig{figure=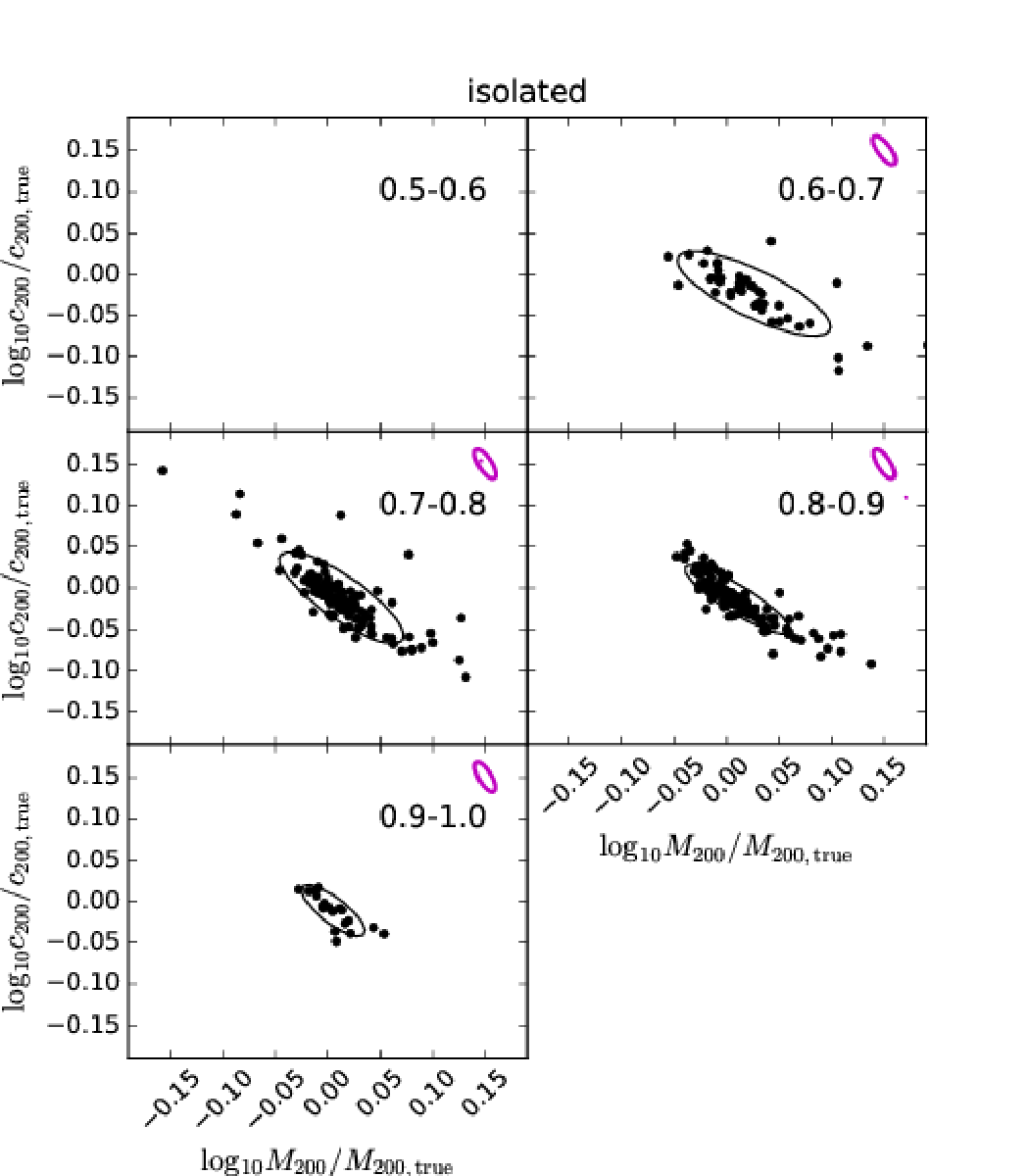,width=0.49\textwidth}
\epsfig{figure=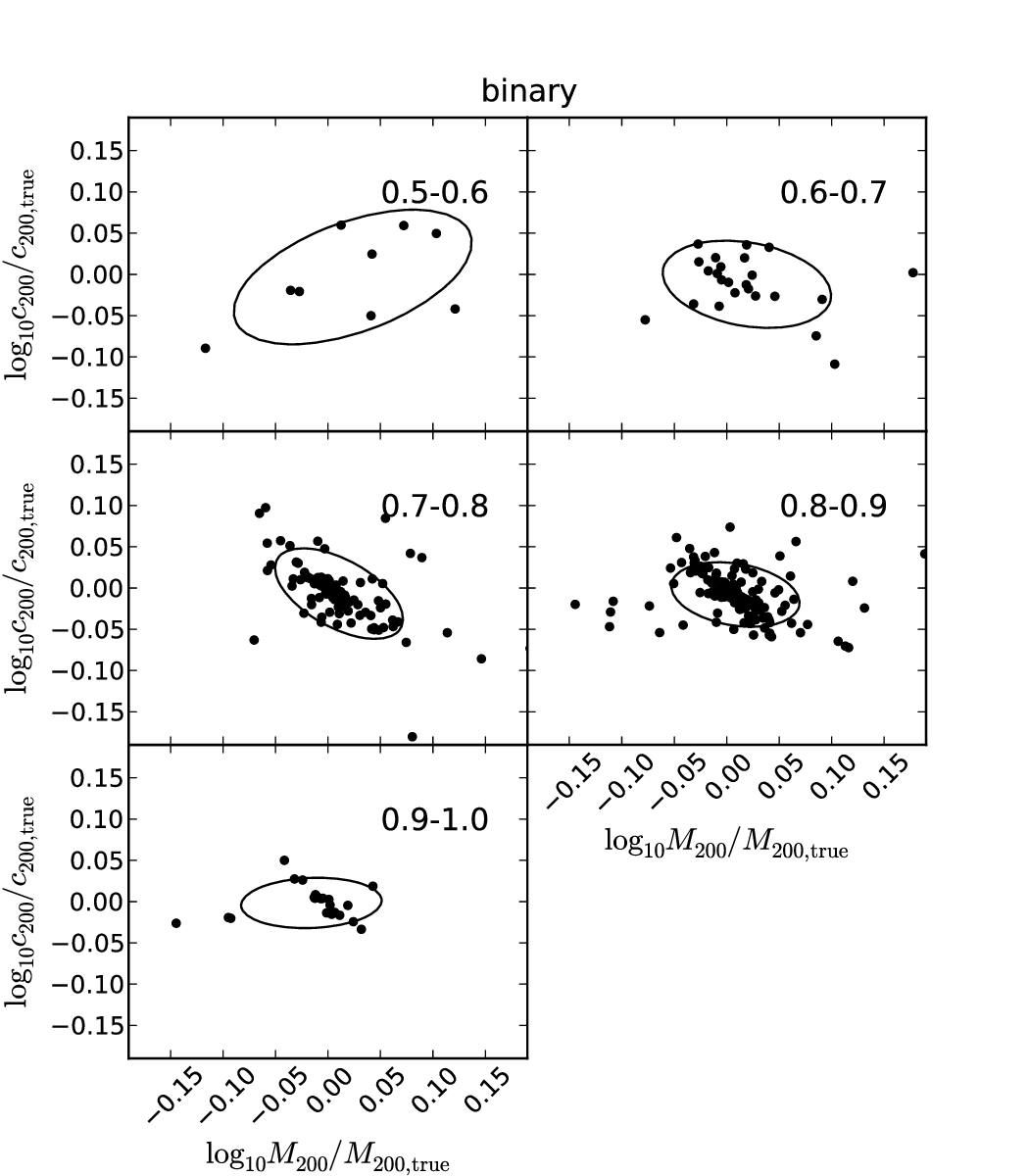,width=0.49\textwidth}
\caption{Concentration and mass parameter values for fits to haloes with different 
minor-to-major axis ratio of the inertial tensor within $R_{200}$ as labelled on 
each panel. Left panels show isolated haloes while right panels show binary haloes.
The little magenta ellipse in each panel of the left plot shows the size of 
the statistical error.}
\label{fig:orient}  
\end{figure*}

Our MRII sample includes both isolated and binary haloes. We have shown results 
based on all these haloes combined in Sec.~\ref{sec:MRII}. Now we analyse the 
two populations separately to see whether there are any systematic differences 
between the two populations. Note \textsc{apostle} haloes are all binaries, and 
thus with our current dataset we are unable to investigate this difference using 
stellar tracers. 

We found the halo mass distribution of binary haloes is biased to be 
smaller than that of isolated haloes. This is because the higher 
abundance of smaller objects enhances the chance of finding them in 
a pair. To avoid possible effects caused by the difference in 
$M_{200}$, we match each of the binary haloes to an isolated halo 
with a similar mass. The criteria is at first chosen to be 
$\Delta \log_{10}M_{200}<$ 0.005 dex and then increased iteratively by 
factors of two upto 0.05 dex. If we fail to find a match with a mass 
difference smaller than 0.05 dex, this halo is discarded. In the end 
we have 332 binary haloes and 332 isolated haloes matched in $M_{200}$.

Results are shown in Fig.~\ref{fig:pairiso}. It is obvious that binary haloes 
have a larger overall scatter in the systematic errors than isolated haloes, 
which is mostly driven by a small fraction of haloes exhibiting large biases 
perpendicular to the anti-correlation direction. 

The fits to dark matter tracers in Aquarius haloes are also shown in 
Fig.~\ref{fig:pairiso}. \cite{han2015b} reported an overall systematic uncertainty
of only 5\% in $M_{200}$ for these haloes, while the systematic uncertainty of MRII 
haloes is about 15\% for isolated ones and 25\% for binaries. Again this 
difference can be attributed to the small size of the Aquarius sample. 
A K-S test between the Aquarius haloes and isolated MRII haloes yields 
a p-value of 0.16 along the anti-correlation direction in the parameter 
space, which indicates an insignificant difference in the distributions.

\subsection{Effect of radial cuts}
\label{sec:cut}
Fig.~\ref{fig:pairiso} can be compared directly with the left panel of Fig.~\ref{fig:pairiso20}. 
The latter adopts a larger inner radius cut of 20~kpc. It is clear that the scatter of $c_{200}$ 
is significantly smaller in Fig.~\ref{fig:pairiso}, whereas the scatter of $M_{200}$ is similar 
in both plots. This is consistent with our findings in Section~\ref{sec:star} that particles 
in the inner region are well relaxed, and the inclusion of them improves the fit. With 
the inclusion of particles within 20~kpc, the effective number of particles is about 5000, in 
contrast to the number of 1000 of Fig.~\ref{fig:pairiso20}.
In particular, the fit to $c_{200}$ is much improved due to the inclusion of tracer particles 
at small radii. This is because $c_{200}$ depends on the scale radius, $r_\mathrm{s}$, which 
is usually about a few to a few tens of kilo-parsecs and can only be well constrained when the 
inner halo is sampled. 

\cite{2015MNRAS.453..377W} have looked at the performance of a distribution function model 
with different outer radius cuts and found tracers within $0.3R_{200}$ give similar best fit 
$M_{200}$ to using all tracers within $R_{200}$ for Aquarius halo A, B, C and D (see their Fig.~17). 
However, once the outer radius cut becomes too small, say, $r<0.3R_{200}$, the bias becomes more 
evident. Moreover, if only tracers within a narrow radial range are used, the result may be 
significantly biased (see their Fig.~16) since significant extrapolations are needed to specify 
the underlying potential in regions where there are no tracers. Though the method tested by 
\cite{2015MNRAS.453..377W} is different, we have also looked at the performance of the 
\textsc{oPDF} by using tracers within $0.3R_{200}$. We found a similar conclusion that 
the overall scatter in the bias remains very similar. This means if only tracers within about 
60 to 70~kpc are available, the result would not be significantly biased from those obtained 
by including tracers in the outskirts. For brevity we do not show results based on the outer 
radius of cut $0.3R_{200}$ as it looks very similar to Fig.~\ref{fig:pairiso}.

\subsection{Spherical symmetry}
\label{sec:shape}

Our current analysis assumes spherical symmetry. In Fig.~\ref{fig:orient}, we test the effect of deviations 
from spherical symmetry on the fits. We split isolated haloes into different 
subsamples according to the minor to major axis ratio, $c/a$, of the inertial 
tensor. The majority of haloes have, in fact, $0.7<c/a<0.9$. There are not 
many haloes with $c/a<0.7$ or $c/a>0.9$. Despite the small number, the contrast 
between haloes with $c/a<0.7$ and $c/a>0.9$ is quite significant, and we see 
a clear trend that the scatter in systematic errors depends on $c/a$.

The typical statistical error size is shown as the magenta ellipse in the 
upper right corner of each panel. The statistical error size is almost independent 
of $c/a$. This means the effective number of independent particles derived in 
Sec.~\ref{sec:MRII} and Sec.~\ref{sec:LG} has been underestimated because the 
systematic uncertainty also has a contribution from violations of spherical 
symmetry. If we choose haloes with $c/a>0.9$, this would allow us to separate 
violations of spherical symmetry and a steady state. For the 
$c/a>0.9$ panel, the systematic uncertainty is about 2.5 times the size of the 
statistical error. The mean effective particle number increased as 
$N_{\rm eff}\sim 18000$. 

The same analysis for binary haloes is shown in the right panels of Fig.~\ref{fig:orient}. 
For each axis ratio bin, the difference between isolated and binary haloes are similar 
to that shown in Fig.~\ref{fig:pairiso}, suggesting that the difference between the two 
is not driven by the difference in halo shapes. Due to the existence in each of the bins 
of highly biased fits perpendicular to the parameter anti-correlation direction, however, 
it is more difficult to observe a clear dependence on $c/a$ for binaries. 

Due to the limited number of \textsc{apostle} haloes and the fact that \textsc{apostle} 
haloes are binaries, the trend is hard to see.

\subsection{Halo merger history}
\label{sec:merger}

\begin{figure} 
\epsfig{figure=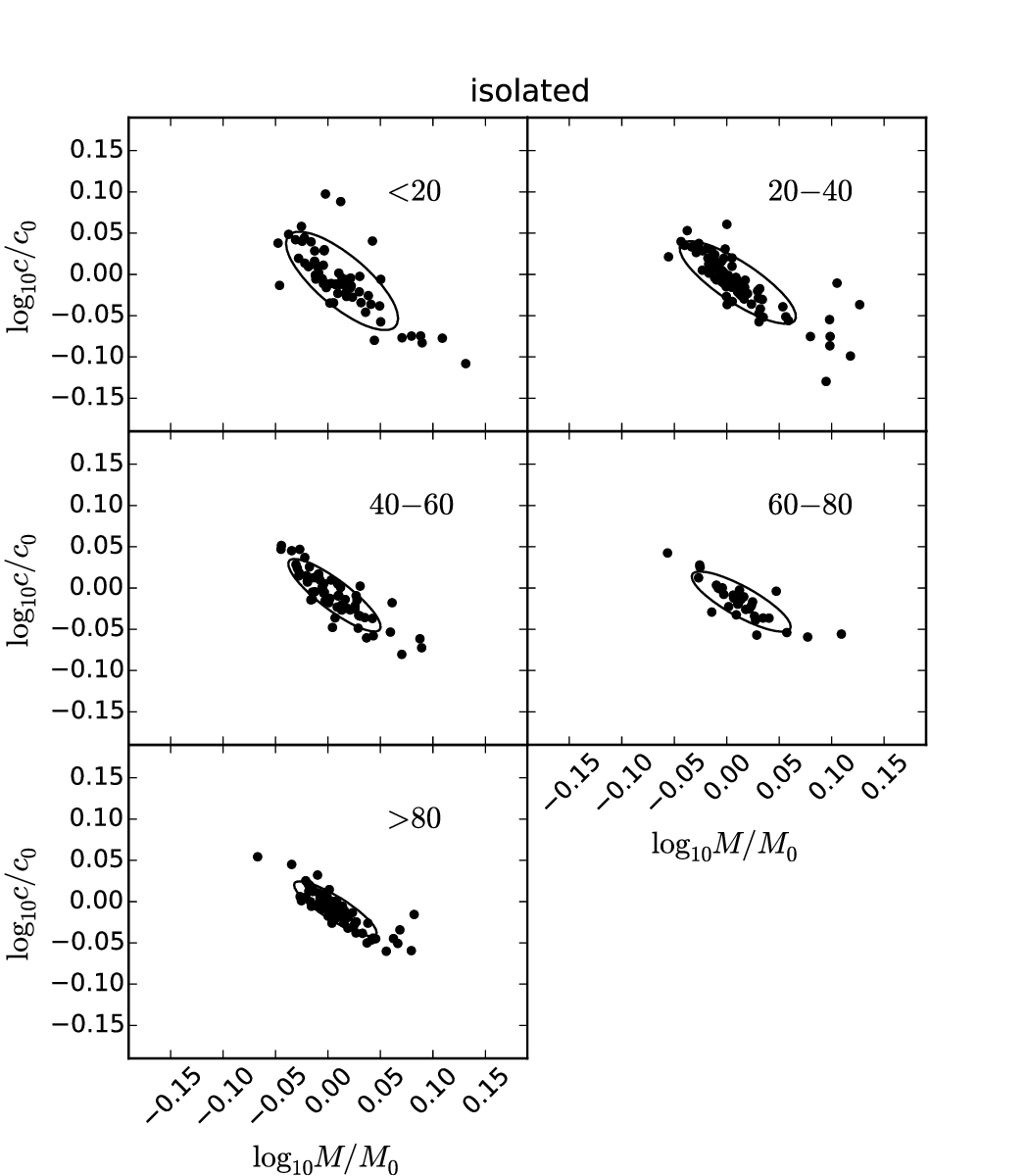,width=0.49\textwidth}
\caption{Concentration and mass parameter values for fits to isolated MRII 
haloes with different $N_{\rm stream,eff}$ defined by Equation~\eqref{eq:Neff}. 
The range of $N_{\rm stream,eff}$ is indicated by the text in each panel.} 
\label{fig:branch} 
\end{figure}

\begin{figure} 
\epsfig{figure=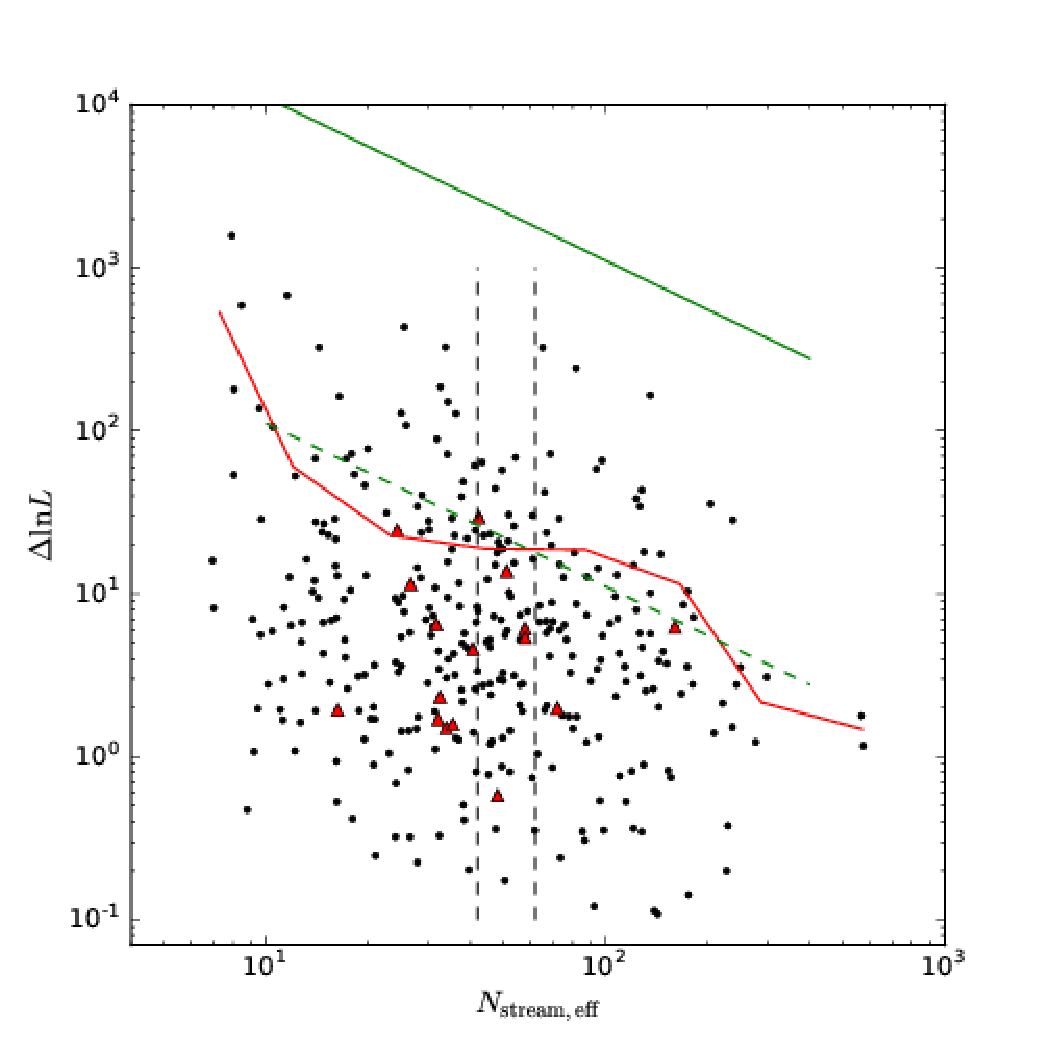,width=0.49\textwidth}
\caption{The log likelihood difference between best fit and true parameters, 
$\Delta \mathrm{ln} L$, versus $N_{\rm stream,eff}$ estimated from Equation~\eqref{eq:Neff}. 
The red solid curve shows the mean $\Delta \mathrm{ln} L$ at fixed $N_{\rm stream,eff}$. 
The green solid line is a theoretical model of $\langle \Delta \mathrm{ln} L(N_{\rm eff})
\rangle=\frac{N}{N_{\rm eff}}$. The green dashed line is obtained with $N_{\rm eff}=100N_{\rm stream,eff}$ 
to match the red solid curve. The median and mean values of $N_{\rm stream,eff}$ are 42 and 
64 respectively, marked by the two vertical black dashed lines. Red triangles mark 
measurements for haloes with $c/a>0.9$.}
\label{fig:branchvslike} 
\end{figure}

We now investigate whether the halo merger history is a hidden variable that 
affects the systematic uncertainty of the fits. To this end, we first seek a quantity 
that characterizes the formation history of the tracer population. The tracer 
particles we use are a mixture of stripped particles from subhaloes/satellites 
and smoothly-accreted particles that did not belong to any bound substructures 
in the past. Particles stripped from the same progenitor are expected to share 
similar orbits and form coherent streams, each of which contains a different 
number of particles. Inside each stream, the particles are highly phase correlated. 
The number of phase-independent particles, $N_{\rm eff}$, is thus determined by 
the number of streams, the size of each stream, and the internal structure of 
them. Neglecting the internal structure, we can derive an effective sample size 
that determines the scatter in the parameter estimate as (see Appendix~\ref{sec:Neff})
\begin{equation}
N_{\rm stream,eff} = \frac{(\sum n_i)^2}{\sum n_i^2},
\label{eq:Neff}
\end{equation} 
where $n_i$ is the number of particles in stream $i$. The summation goes over all 
the phase-space structures including smoothly-accreted particles. It is easy to 
prove that $1\leq N_{\rm stream,eff} \leq m$, where $m$ is the number of streams. 
$N_{\rm stream,eff}$ is smallest when the sample is dominated by a single stream 
($n_1\sim N$ and $n_{i\neq 1}\sim 0$), while it is largest when the streams are 
of equal size. If the particles inside each stream share a common phase-space 
coordinate, we expect $N_{\rm eff}=N_{\rm stream,eff}$. In reality, however, this 
assumption does not hold and we expect $N_{\rm stream,eff}$ to be only a crude 
estimate of $N_{\rm eff}$. We will further discuss their relation later in this section.

$N_{\rm stream,eff}$ is closely related to the merger history of haloes. 
To determine $N_{\rm stream,eff}$, we trace the particles in our sample back 
to their progenitors. Particles stripped from the same progenitor are considered 
to be within the same phase-space structure, while each smoothly-accreted 
particle is treated as an independent phase-space structure containing only 
one particle. Note we trace the particles back hierarchically, that is, 
if a particle merges from a small halo to a large halo, and then the large
halo merges with the final host, particles originally belonging to the small halo 
will be assigned the small halo as their progenitor. $N_{\rm stream,eff}$ can 
then be estimated for each halo and can be used to characterize the merger 
history. 

The results are shown in Fig.~\ref{fig:branch} for isolated MRII haloes. 
We only focus on isolated objects because as we have seen, the trend is harder 
to see for binaries. It is encouraging to see
a weak but significant trend for haloes with larger 
$N_{\rm stream,eff}$ to exhibit smaller scatter.
Similar trends also exist for stellar tracers in \textsc{apostle} haloes.

A more obvious trend is revealed in Fig.~\ref{fig:branchvslike}. The log 
likelihood difference, $\Delta \mathrm{ln} L$, reflects the level of systematic 
uncertainties.\footnote{Or more precisely, it is a measure of the signal to noise 
of the systematic uncertainties, $2\Delta \ln L=||{b}/{\sigma}||^2$.} From a statistical point 
of view, it is important to realize that $\Delta \ln L$ is a random variable 
resulting from fitting one random realization of the underlying model.
If the data are generated from the model, then according to Wilks' 
theorem~\citep{Wilks}, the log-likelihood ratio $2\Delta \ln L$ follows 
a $\chi^2(n)$ distribution with $n$ degrees of freedom, where $n$ is the 
number of free parameters in the fit. In our case, we expect $2\Delta \ln L$ 
to behave like a $\chi^2(2)$ variable if the tracers follow the steady-state 
distribution in each halo and if there is no violation of spherical 
symmetry. In the presence of phase correlations, the likelihood 
ratio resulting from the phase-independent particles is still a $\chi^2(2)$ 
variable, while that of the full sample would behave like a scaled variable 
$({N}/{N_{\rm eff}})\chi^2(2)$, where $N$ is the sample size and 
$N_{\rm eff}$ is the number of phase-independent particles. In particular, 
we expect 
\begin{equation}
\avg{\Delta\ln L}=\frac{1}{2}\frac{N}{N_{\rm eff}} \avg{\chi^2(2)}
=\frac{N}{N_{\rm eff}},\label{eq:lnL_N}
\end{equation}
if ignoring the violation of spherical symmetry.

If our $N_{\rm stream, eff}$ is a correct estimate of $N_{\rm eff}$, we 
should see a dependence of $\Delta\ln L$ on $N_{\rm stream,eff}$. Fig.~\ref{fig:branchvslike} 
shows there is indeed a trend of decreasing $\Delta \mathrm{ln} L$ with 
increasing $N_{\rm stream,eff}$. The scatter of $\Delta \mathrm{ln} L$ 
is quite large, which is expected if $\Delta \ln L$ is a scaled $\chi^2$ variable. 
More quantitatively, $\avg{\Delta\ln L}$ scales with the inverse of $N_{\rm stream,eff}$, 
consistent with the expectation from Equation~\eqref{eq:lnL_N}. However, there 
is an offset between the expected $\avg{\Delta\ln L}$--$N_{\rm eff}$ relation 
and the actual $\avg{\Delta \ln L}$--$N_{\rm stream,eff}$ relation. This 
reflects that our $N_{\rm stream,eff}$ on average underestimates $N_{\rm eff}$, 
i.e., $N_{\rm eff}=\alpha N_{\rm stream, eff}$, with $\alpha\approx100$. 
For haloes with $c/a>0.9$ (red triangles in the figure), $\Delta\ln L$ 
is systematically smaller due to the decrease in systematic errors, and thus 
$\alpha\approx400$ for the most spherical haloes.

This correction factor is easy to understand because we have ignored the internal 
structure of the streams when deriving $N_{\rm stream,eff}$, while in reality 
the particles inside the streams are not completely correlated with each other 
and could contribute an additional number of phase-independent particles. 
In addition, each progenitor could have given rise to multiple streams, 
further increasing the number of phase-independent particles. This is not 
considered in Equation~\ref{eq:Neff}. Despite this underestimation, 
$N_{\rm stream,eff}$ correctly captures the variation of the likelihood 
ratio (or uncertainty level).

The average $N_{\rm eff}$ derived from $\alpha N_{\rm stream,eff}$ is about $5000$ 
for all haloes and about $20000$ for haloes with $c/a>0.9$. The same factor 
is obtained by comparing the statistical and systematic uncertainties for 
isolated haloes in Fig.~\ref{fig:pairiso} and for haloes in the last panel 
of Fig.~\ref{fig:orient}.

\section{Discussion}
\subsection{Understanding the parameter correlations}
\label{sec:correlation}

\begin{figure*} 
\epsfig{figure=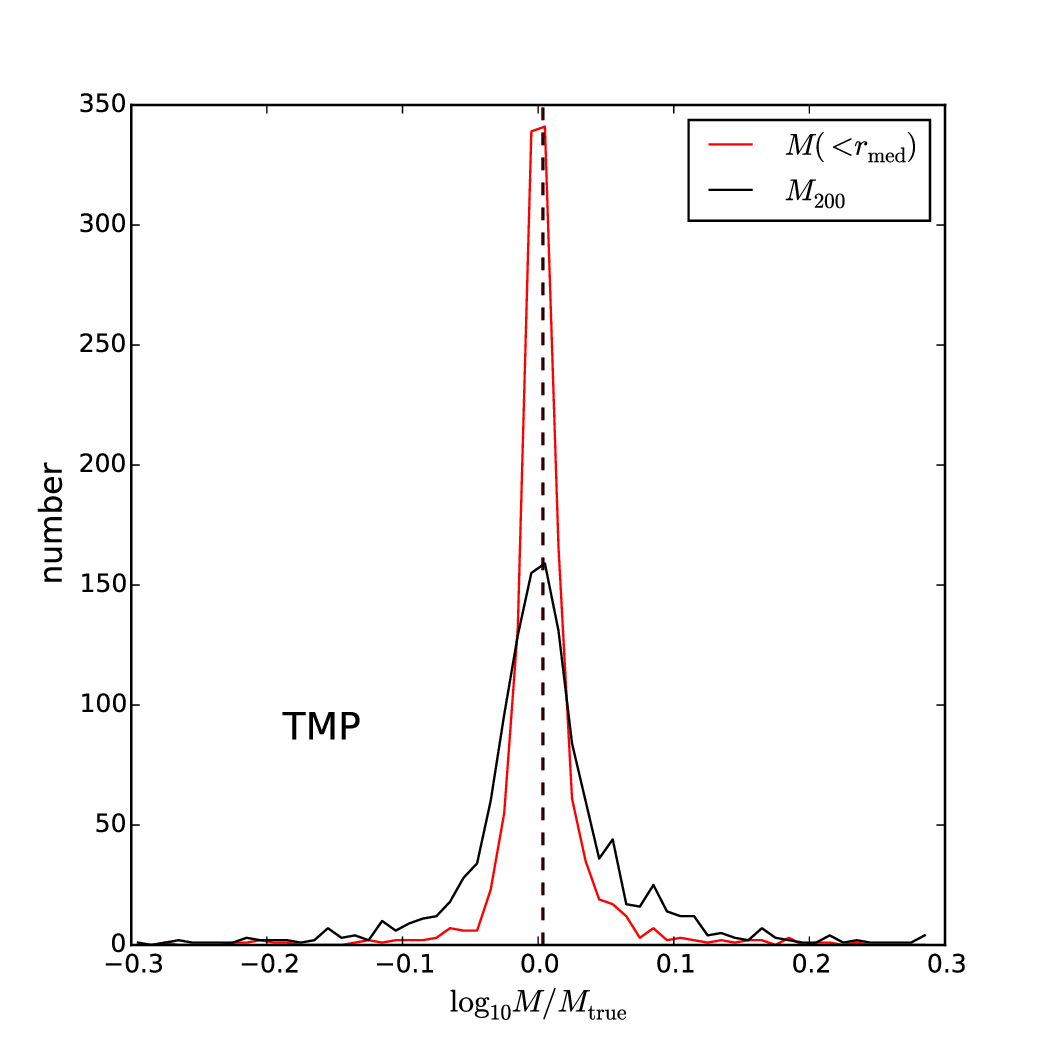,width=0.49\textwidth}
\epsfig{figure=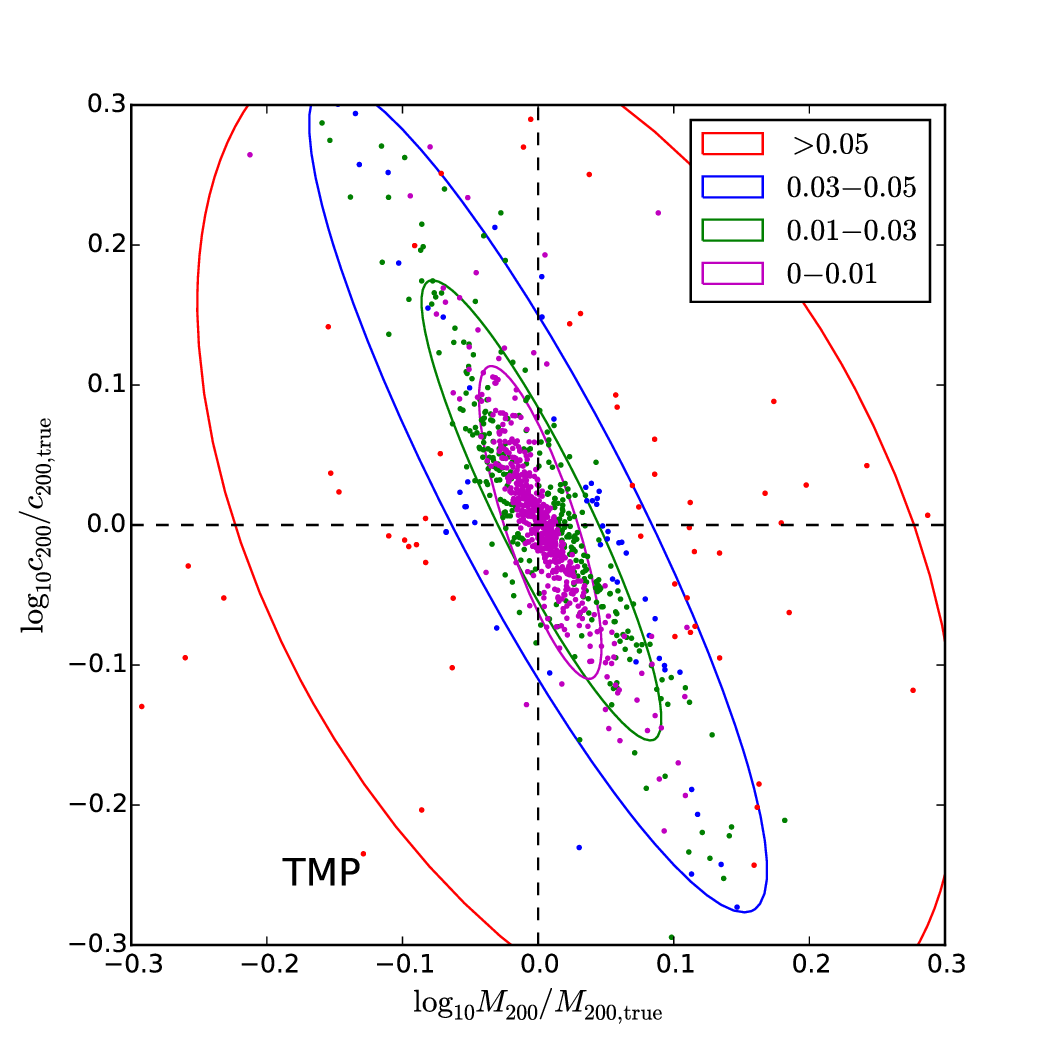,width=0.49\textwidth}
\caption{{\bf Left:} The distributions of the fitted $M_{200}$ (black) and 
the fitted mass within median radius of tracers (red), in units of 
their true values. Vertical dashed lines with the corresponding colours mark 
the average values over the two distributions respectively. {\bf Right:} 
Similar to the panels in Fig.~\ref{fig:pairiso20}, but the measurements 
are colour coded by the values of $\log_{10}M(<r_\mathrm{med})/M(<r_\mathrm{med})_\mathrm{true}$ 
as labelled in the legend.} 
\label{fig:halfdependence} 
\end{figure*}

The anti-correlation between mass and concentration parameters is commonly seen in dynamical modelling 
of the galactic potential \citep[e.g.][]{2014MNRAS.439.2678D,2014ApJ...794...59K,2015MNRAS.453..377W}. 
Despite the difference to other methods, we also see a strong anti-correlation between $M_{200}$ and 
$c_{200}$. What is the reason behind the parameter correlation?

This can be understood intuitively in the following way. The fundamental quantity constrained by 
the observed dynamics of tracer particles is essentially the rotation curve of the system, 
$V_{\rm circ}^2(r)={\rm G}M(r)/r$. For any tracer population, we expect the rotation curve to be 
best constrained near a characteristic radius $r_{\rm c}$ where most of the tracer particles are 
located. A natural choice of $r_{\rm c}$ would be the median radius of the tracer (although not 
exactly, see~\citealp{han2015a}). Equivalently, the mass inside the characteristic radius, 
$M(r_{\rm c})$, is well constrained, as we demonstrate explicitly in Appendix~\ref{sec:prof}. 
As a result, any mass distribution allowed by the observed dynamics of the tracers has to cross 
the $M(r_{\rm c})$ point in the mass profile, which has already been discussed extensively 
by \cite{han2015a}, who also discussed differences to other studies 
\citep[e.g.][]{2009ApJ...704.1274W,2010MNRAS.406.1220W,2011MNRAS.411.2118A}. This leads to a 
tight correlation between the amplitude and slope of the mass profile near the virial radius. 
Since a shallower slope roughly corresponds to a larger concentration, the mass-slope correlation 
translates into the anti-correlation between the mass and concentration parameters. 

Now it can also be understood that the location of the characteristic radius determines the amount 
of correlation between the parameters: the closer $r_{\rm c}$ is to $R_{200}$, the weaker the 
correlation. In other words, the shape of the covariance ellipse in the parameter plane is 
determined by the location of $r_{\rm c}$, or the radial distribution of the tracer, as explicitly 
demonstrated by \citet{han2015a}. Since dark matter is more extended than stars, the median 
radius of dark matter is closer to $R_{200}$ than that for stars. We would expect a weaker parameter 
correlation for dark matter than stars as tracers, which we discuss more in the Appendix~\ref{sec:prof}. 

We summarise the model performance in recovering the mass inside the median radius 
of the tracer, $M(<r_\mathrm{med})$, in the left panel of Fig.~\ref{fig:halfdependence}, 
where the distribution of ratios between best fit and true masses are plotted. It is 
clear that the fits of $M(<r_{\rm med})$ show less scatter from the true values than 
those of $M_{200}$, while both types of fit are ensemble unbiased. In 
Appendix~\ref{sec:prof}, we show how the mass profiles, $M(<r)$, are constrained over 
the whole radial range using \textsc{apostle} haloes.

The right plot of Fig.~\ref{fig:halfdependence} shows that the uncertainty
in $M(<r_\mathrm{med})$ is closely related to where the measurements sit in 
the ($M_{200}, c_{200}$) parameter plane. It is clear that measurements with 
larger values of $\log_{10} M(<r_\mathrm{med})/M(<r_\mathrm{med})_\mathrm{true}$ 
have larger scatter and deviate more from the parameter anti-correlation 
direction. Equivalently, haloes whose best fit parameters deviate away from 
the anti-correlation are also those that have a biased fit in $M(<r_\mathrm{med})$.

\subsection{Effect of modelling the potential with an NFW profile}
\label{sec:nfw}

\begin{figure*} 
\epsfig{figure=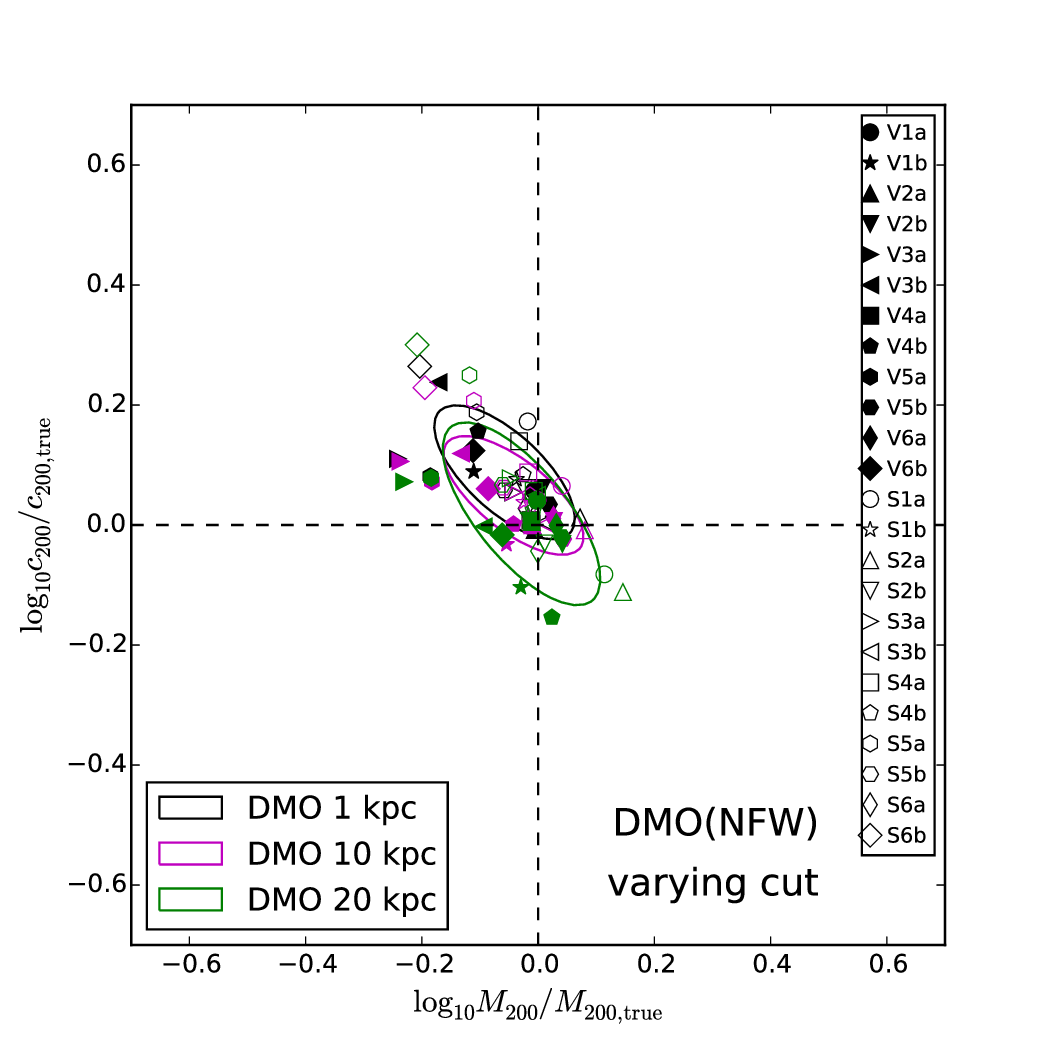,width=0.49\textwidth}%
\epsfig{figure=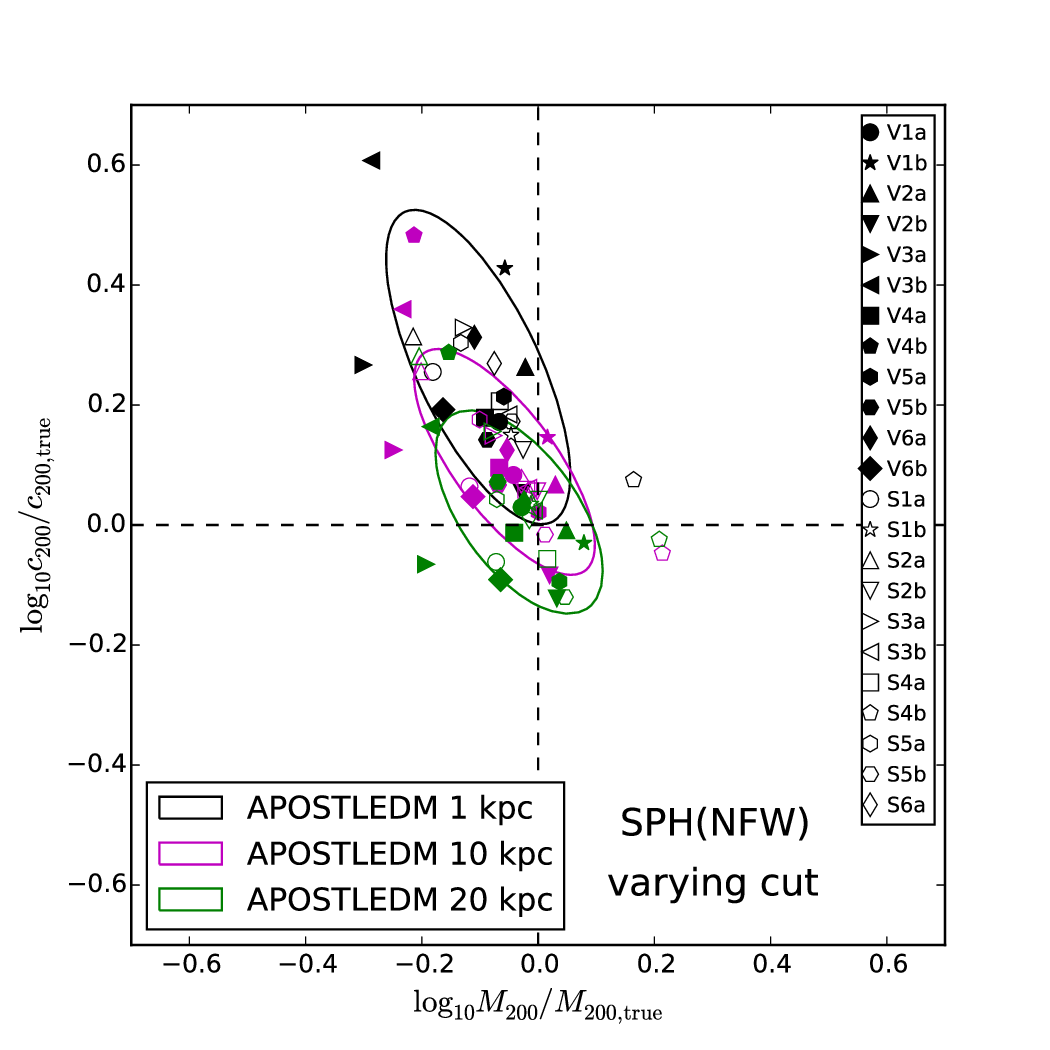,width=0.49\textwidth}%
\caption{{\bf Left:} Fits to \textsc{apostle} haloes using DM particles as tracers. The underlying 
potential is parametrized in an NFW form. Black, magenta and green symbols (ellipses) refer 
to inner radius cuts of 1~kpc, 10~kpc and 20~kpc, respectively. Ellipses mark the $1~\sigma$ 
scatters of the fitted parameters.  The left and right panels are for the DM-only (DMO) and 
hydrodynamical (SPH) runs respectively.} 
\label{fig:dmnfw} 
\end{figure*}

In all the previous sections the underlying potential profiles are modelled 
using potential templates, which are guaranteed to match the true potential when using the 
true parameters. However, for haloes in the real Universe we do not know the true potential 
a priori, hence such templates are unavailable. Instead, one has to assume some practical 
parametrizations of the profile. Deviations of the true profile from the assumed functional 
form could introduce additional uncertainty to the model fits. In this subsection we 
study this source of uncertainty focusing on the commonly adopted NFW parametrization.

Fig.~\ref{fig:dmnfw} shows the the fit to \textsc{apostle} haloes using DM particles as tracers, 
adopting different inner radius cuts. First of all, we note for a fixed halo, the best fit 
halo parameters are not the same between the SPH and the corresponding DMO runs. This is 
because particles between the two simulations are not expected to match in phase space, 
due to the implementation of baryonic physics and their non-linear orbital evolution. 

The fits to DMO haloes appear largely unbiased on average, although there is a weak 
tendency of an overall uncertainty when adopting smaller radial cuts. This is consistent with the 
fits in Fig.~\ref{fig:pairiso20} and Fig.~\ref{fig:pairiso} when adopting template profiles. 
This means the NFW parametrization is a reasonably good model for the underlying potential 
of haloes in DMO simulations, consistent with the findings in \cite{han2015b} using 
Aquarius simulations. We have checked that the conclusion holds with the much larger 
sample of MRII haloes as well. However, for SPH runs, the fits become more and more biased 
as the inner cut becomes smaller and smaller. This suggests that the deviation from NFW 
parametrization is much larger in the inner halo for SPH runs. Note we have explicitly 
confirmed that such an increase of the bias with decreasing inner radius cut is not 
present when the true potential templates are used. Indeed, using the \textsc{eagle} simulation 
\cite{2015MNRAS.451.1247S} found that the presence of stars can produce cuspier inner 
profiles than the NFW model, and the effect is most prominent in haloes of masses about 
$10^{12}$ to $10^{13}\mathrm{M_\odot}$. We have checked that \textsc{apostle} haloes do 
indeed deviate more from NFW than those in the DMO runs. In the Appendix, we explicitly 
compare the best fit to the true halo density profiles of \textsc{apostle} haloes. 

We have repeated the left plot of Fig.~\ref{fig:dmnfw} using star particles from 
\textsc{apostle} as tracers, and we found similar trends that as inner radius 
cuts are reduced the best fit parameters using the NFW model are biased 
more towards high concentration and low mass, while the scatter remains almost 
unchanged. A comparison with Fig.~\ref{fig:insitu} reveals that the improvement 
due to the inclusion of in-situ stars can only be achieved if the underlying 
potential is properly modelled. 

Finally, we note that the scatter in $c_{200}$ is at best only slightly decreased
in Fig.~\ref{fig:dmnfw} when the inner radius cut is decreased, in contrast 
to the significant reduction from Fig.~\ref{fig:pairiso20} to Fig.~\ref{fig:pairiso}. 
This can be understood because the constraint on $c_{200}$ can only be significantly 
improved if the inner potential profiles are correctly modelled. 

\section{Conclusions}
\label{sec:concl}

We have studied the dynamical state of a large sample of MW size haloes using a 
general dynamical method, \textsc{oPDF}, that depends on a minimum number of 
assumptions, namely the time-independence of the distribution function and the 
spherical symmetry of the halo potential. Because these two assumptions are 
often adopted in many other dynamical models, our analysis can be used 
to understand the minimum amount of uncertainty in these models arising 
from the two assumptions. 

The tracers used include dark matter particles in isolated and binary haloes 
from the Millennium~II (MRII) simulation, as well as stars and dark matter 
particles in 24 haloes (or 12 halo pairs) from the \textsc{apostle} hydrodynamical 
simulations. For direct comparisons we also applied \textsc{oPDF} to dark 
matter particles in corresponding dark matter only runs of \textsc{apostle}. Binary 
haloes in MRII and \textsc{apostle} halo pairs are selected in analogy to the MW-M31 
pair. The large sample of haloes have enabled us to thoroughly test how 
\textsc{oPDF} works in recovering the halo potential, for haloes with different 
properties. We model the underlying potential profiles using parametrized 
templates generalised from the true potential. This enables us to separate 
the effect of potential parametrization from other factors in the modelling. 
In addition, we also tried modelling the potential profiles using NFW profiles, 
which helps us to quantify the effect of a practical parametrization 
on the modelling.
  
For each halo, we fit for the mass and concentration parameters of the halo. 
We find the fit to each individual halo is biased in a stochastic way, with 
a large halo-to-halo scatter. The scatter is larger than those previously 
reported by \cite{han2015b} using the smaller sample of Aquarius haloes. 
Adopting an inner radius cut of 20~kpc for tracers, we found the scatter 
can be as large as a factor of three if stars are used as tracers. Dark 
matter particles, on the other hand, give much smaller scatter of about 
25\% for $M_{200}$ and 25\% to 40\% for $c_{200}$. The scatter in $c_{200}$ 
can be reduced to less than 25\% if including dark matter particles in 
the very central region with proper modelling of the inner density or 
potential profiles. This is because $c_{200}$ depends sensitively on 
$r_s$ and hence on tracers in the very inner region.

The large scatter in best fit halo parameters based on stars as tracers 
is worrying. On the one hand, we should be cautious about possible model 
dependencies, since the reliability of the systematic error size 
depends on how realistically the hydrodynamical simulation reflect 
the real world. On the other hand, if we assume the hydrodynamical 
simulations are realistic enough, the amount of systematic scatter would 
have practical implications for dynamical modelling of galactic haloes. 
The systematic errors we see with \textsc{oPDF} should also exist and 
be irreducible for any dynamical models that also make the steady 
state and spherical assumptions. Previous studies challenging 
the $\Lambda$CDM cosmological model with the abundance and dynamics of 
dwarf satellite galaxies often depend sensitively on the mass of the
MW \citep[e.g.][]{2011MNRAS.415L..40B,2012MNRAS.422.1203B}. 
If the MW halo mass is reduced by a factor of two, the challenge 
would no longer exist \citep{2012MNRAS.424.2715W}. {\it Given the 
large uncertainties behind dynamical models, it is not surprising 
to expect a factor of two uncertainty in the measured MW halo 
mass. It is thus dangerous to draw strong conclusions based on any 
single value of the MW halo mass without quoting the uncertainties. }

With the much larger sample of haloes, we found if the underlying potential 
profiles are correctly modelled, the best fit halo parameters averaged over 
different haloes are {\it ensemble unbiased}, despite the large scatter from 
halo to halo. In addition, the correlation of the systematic uncertainties
tend to be aligned with the statistical noise, which is controlled to be 
much smaller than the typical systematic uncertainty observed. 
\cite{2015MNRAS.453..377W} and \cite{han2015a} suggest the explanation is 
that the statistical errors have been underestimated due to the phase-space
correlations of particles in streams. Since statistical errors are expected 
to be ensemble unbiased, this supports our hypothesis that the statistical 
errors are underestimated.


The systematic scatter, despite being large, tends to happen along 
a direction of anti-correlation between $M_{200}$ and $c_{200}$. 
\cite{2015MNRAS.453..377W} and \cite{han2015a} have discussed that 
the anti-correlation reflects a fundamental quantity, the total mass 
enclosed within a certain characteristic radius, which can be constrained 
more robustly than the mass or concentration parameters. The characteristic 
radius is very close to the median radius of tracers. We revisited the 
conclusion using our much larger samples of haloes and reached the same 
conclusion. We found haloes whose mass within the median tracer radius 
cannot be well constrained usually have large uncertainties in 
their best fit halo parameters or the best fit parameters deviate away 
from the anti-correlation direction.

If the underlying potential is not correctly modelled, the ensemble 
averaged best fit parameters can be biased. For \textsc{apostle} haloes, once 
we include particles in the very inner region ($r<20$~kpc) as tracers 
and model the underlying potential using the NFW model, the best fit 
parameters become systematically overestimated in $c_{200}$ and 
underestimated in $M_{200}$. This is because the NFW model fails to 
properly describe the density profiles in the very inner region. For
DM only simulations, the effect is negligible, however, which warns 
against the use of pure $N$-body simulations in such studies. 

Comparing isolated haloes with binaries, we found the population 
of binary haloes tend to have a larger scatter in the best fit halo 
properties. Since binary haloes stay in dense regions, we expect their 
dynamics are perturbed by not only the massive companion in the 
pair but also the rich population of smaller companions or substructures. 
However, we need to be cautious about interpreting the results, 
because they are based on dark matter particles as tracers, which are 
more extended than stars and could be affected more strongly 
by nearby companions. 

Looking at possible dependencies of the systematic uncertainty on various halo 
properties, we fail to detect any dependence on halo mass or concentration. 
For binaries, we do not detect dependence on pair separation or mass 
ratio. This might be due to the limited dynamic range of these properties 
in our sample. For isolated haloes, we observe a clear dependence of 
the uncertainty on the minor to major axis ratio of the inertial tensor 
of the halo, reflecting the effect of deviations from spherical symmetry. 
There is a significant dependence of the scatter of biases on the 
(weighted) number of phase-space structures, which is closely related 
to the halo {\it merger history} and the number of {\it phase-independent 
particles}. This directly supports our interpretation that {\it the uncertainty}
is related to the number of independent phase-space structures in the 
halo.

Assuming the statistical and systematic errors would have comparable 
size once we properly consider the true degree of freedom contributed by 
independent particles, we can make crude estimates of the effective number 
of these independent particles, of about 40 for halo star particles 
in \textsc{apostle} haloes and 1000 for dark matter particles beyond 20~kpc 
in MRII haloes. The larger number of effective particles for dark matter 
tracers is consistent with the picture that dark matter particles are more 
phase mixed and relaxed. However, since the systematic uncertainty is 
also determined by violations of the spherical assumption, this effective 
number of independent particles would be increased by about 
a factor of 4 for the most spherical haloes.

These numbers are related to the (weighted) number 
of phase-space structures and reflect the intrinsic dynamical state of 
the halo. {\it They have important and useful implications for real observations: 
the uncertainty in the dynamical inference saturates to an intrinsic 
uncertainty determined by the dynamical state of the halo, once the 
real sample size becomes much larger than the effective sample size. 
Further increasing the sample size beyond that does not help in reducing 
the uncertainty in the estimates of halo properties.}

\section*{Acknowledgements}
WW is grateful for useful discussions of merger trees with John Helly 
and Yan Qu. WW also thank Azadeh Fattahi and Shi Shao about the useful 
information of \textsc{apostle}. This work was supported by the European Research 
Council [GA 267291] COSMIWAY and Science and Technology Facilities 
Council Durham Consolidated Grant[ST/F001166/1,ST/L00075X/1]. WW 
acknowledges a Durham Junior Research Fellowship (RF040353). The 
simulations for the Aquarius project were carried out at the Leibniz 
Computing Centre, Garching, Germany, at the Computing Centre of the 
Max-Planck-Society in Garching, at the Institute for Computational 
Cosmology in Durham, and on the STELLA supercomputer of the LOFAR 
experiment at the University of Groningen. \textsc{apostle} project used the 
DiRAC Data Centric system at Durham University, operated by the 
Institute for Computational Cosmology on behalf of the STFC DiRAC 
HPC Facility (www.dirac.ac.uk), and also resources provided by 
WestGrid (www.westgrid.ca) and Compute Canada (www.computecanada.ca).
The DiRAC system was funded by BIS National E-infrastructure capital 
grant ST/K00042X/1, STFC capital grants ST/H008519/1 and ST/K00087X/1, 
STFC DiRAC Operations grant ST/K003267/1 and Durham University. 
DiRAC is part of the National E-Infrastructure. Kavli IPMU was 
established by World Premier International Research Center Initiative 
(WPI), MEXT, Japan. This work was supported by JSPS KAKENHI Grant 
Number JP17K14271.

\bibliography{master}

\appendix
\section{The fitted mass profiles}
\label{sec:prof}

\begin{figure*} 
\epsfig{figure=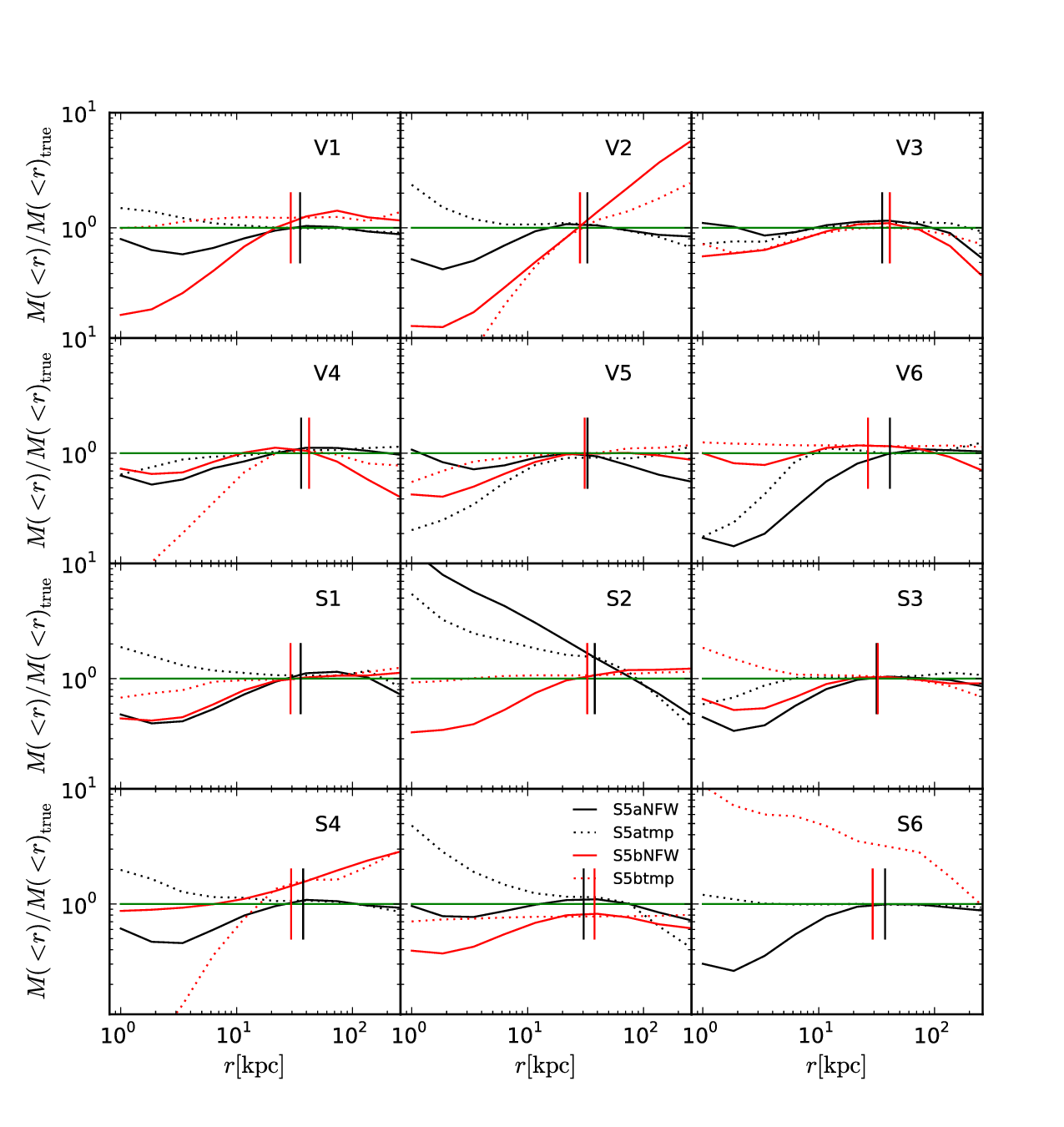,width=0.9\textwidth}%
\caption{The best fit versus true cumulative mass profiles of \textsc{apostle} 
haloes. star particles are used as tracers with an inner radius cut of 20 
kpc. Each panel refers to one \textsc{apostle} simulation volume, as indicated 
by the label. In each panel, black and red lines refer to the MW and M31 
analogues. Solid and dashed lines are based on best fit profiles obtained 
using the NFW profile and true potential templates, respectively. Solid 
vertical lines mark the position of median radius of stellar tracers 
in the two haloes. The green horizontal line marks where the best fit 
and true profiles agree. The red solid curve is missing in the S6 panel, 
because halo S6b is extremely perturbed and the fit fails to converge. } 
\label{fig:profstar} 
\end{figure*}

\begin{figure} 
\epsfig{figure=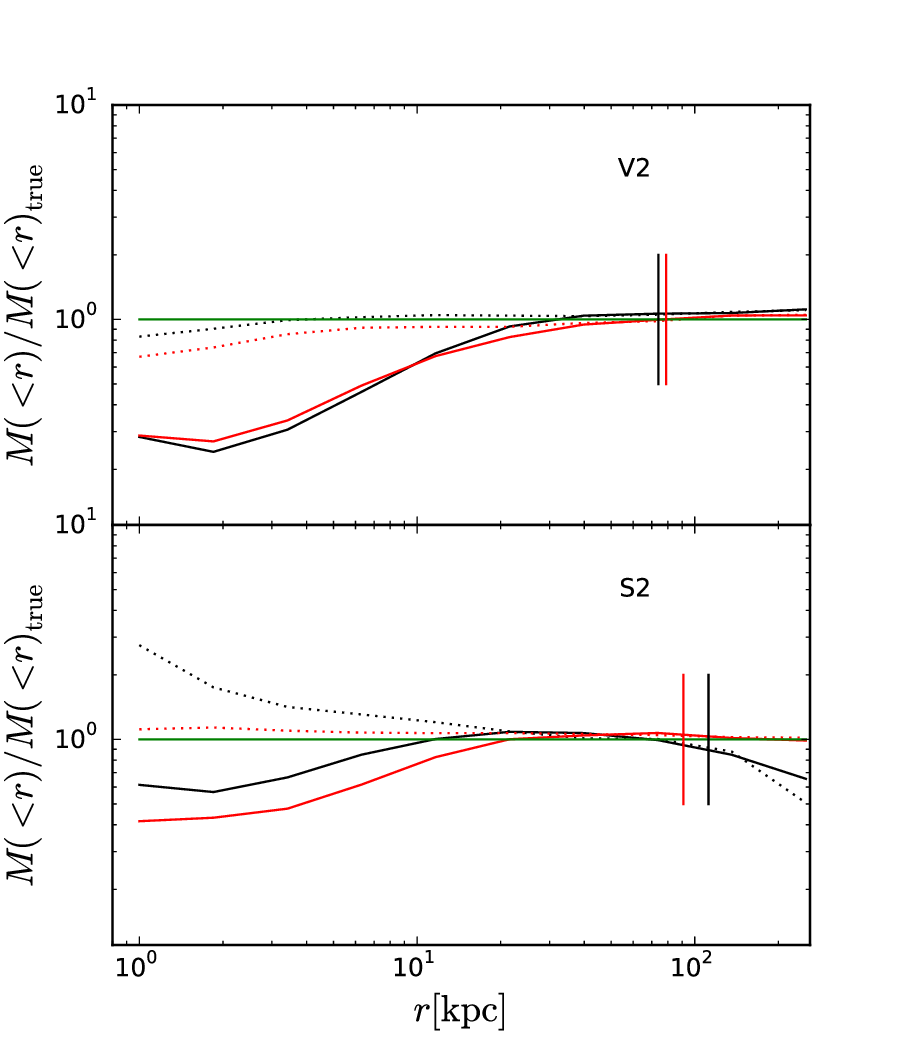,width=0.49\textwidth}%
\caption{As Fig.~\ref{fig:profstar}, but is based on dark matter particles 
in \textsc{apostle} haloes as tracers. An inner radius cut of 20 kpc has been 
adopted. For brevity we only shown results for simulation volumes V2 and 
S2.} 
\label{fig:profdm} 
\end{figure}

We have seen in Sec.~\ref{sec:correlation} that the total mass within 
the median radius of tracers can be constrained much better than 
the virial mass, $M_{200}$. In this appendix we not only look at the 
mass within a fixed radius, but also investigate how well the whole 
mass profile is recovered for each halo. We present results 
based on both star particles and dark matter tracers in \textsc{apostle}.  

Fig.~\ref{fig:profstar} shows the result using stars as tracers. It 
is clear that the mass profile is recovered best near $r_{\rm med}$, 
while it can be biased at both larger and smaller radii. In 
particular, the fit almost always leads to a large bias in the 
inner mass profile when an NFW parametrization is adopted, even 
though in many cases the outer mass profile can still be recovered well. 
This directly confirms our conclusion in Sec.~\ref{sec:nfw} 
that the NFW model fails to properly model the inner density 
profiles.  

Fig.~\ref{fig:profdm} is similar to Fig.~\ref{fig:profstar}, but we 
use dark matter particles in \textsc{apostle} haloes as tracers. For brevity 
we only show results for volumes V2 and S2, as the conclusions based on 
the other haloes are the same. \cite{han2015b} has concluded that 
dark matter particles as tracers can give better constraints on 
$M_{200}$ than star particles. Comparing Fig.~\ref{fig:profdm} with 
Fig.~\ref{fig:profstar}, it is obvious that the mass at all radii 
can also be better constrained.

Since for a fixed halo the underlying potential is the same regardless of 
what kind of tracer particles are used, the better constraints when using 
dark matter particles rather than stars as tracers 
cannot be due to violations 
of the spherical assumption. It is, however, possible that dark matter 
particles are more extended than stars and thus probe better the underlying 
potential in the outskirts. Following \cite{han2015b}, we have picked 
subsamples of dark matter particles having the same binding energy and 
angular momentum distributions as stars. The radial distributions of 
these dark matter particles are as concentrated as the stars, but 
result in a similar amount of scatter in systematic errors
as for the full dark matter samples. Thus the main reason 
for the difference is that dark matter particles are more dynamically 
relaxed than halo stars. This confirms the conclusion of \cite{han2015b} 
is true not only for the mock stellar halo catalogue created 
by the particle tagging approach, but also for star particles 
in hydrodynamical simulations.

The median tracer radius is a few tens of kilo parsec for stars, while 
it is much closer to 100~kpc for dark matter, i.e., two to three times 
larger. We have discussed in Sec.~\ref{sec:correlation} that if the 
mass within a certain radius that is much larger than the characteristic 
radius of the tracer population is chosen as a free parameter, it 
will be strongly correlated with the shape or concentration. Since 
the median radius of dark matter particles is closer to $R_{200}$, 
we expect the anti-correlation between $M_{200}$ and $c_{200}$ (see 
Sec.~\ref{sec:correlation}) to be weaker when using dark matter 
particles as tracers. We have checked and found the normalised 
covariance between $M_{200}$ and $c_{200}$ is very close to 1 when 
star particles are used as tracers. The covariance indeed becomes 
smaller (about 0.6 to 0.8) for dark matter particles, but is still 
strong because the covariance coefficients are mostly above 0.5. 
However, the better constraint on $M_{200}$ from using dark matter 
as tracers is unlikely to be mainly explained by the weaker 
parameter correlation, given the fact explained above that subsamples 
of dark matter particles that have the same radial distribution as 
stars show similar scatter in the best fit halo parameters.  

\section{The effective sample size}
\label{sec:Neff}
To derive the effective sample size, it is simplest to start from 
the mean phase estimator~\citep{han2015a} of the halo parameters. 
The radial phase angle of tracer particles in a steady state system has
a uniform distribution. For a sample of $N$ particles drawn from a uniform 
phase distribution, the mean radial phase angle, $\bar{\theta}$, is expected 
to follow a normal distribution with mean 0.5 and standard deviation $1/\sqrt{12N}$. 
The normalized mean phase, $\bar{\Theta}=\sqrt{12N}(\bar{\theta}-0.5)$, can 
then be used as a measure of the deviation of the actual phase distribution 
from the expected uniform distribution. If the data agree with the model, 
$\bar{\Theta}^2$ from different realisations of the same distribution follows 
a $\chi^2$ distribution with one degree of freedom. Hence, one can quantify 
the discrepancy level of the data from the model through the probability of 
obtaining a value of $\chi^2$ as extreme as the measured value of $\bar{\Theta}^2$.

\cite{han2015a} found that the confidence interval of the likelihood 
estimator is comparable with that of the mean phase estimator except 
for a parameter degeneracy in the latter. As a result, we can estimate 
the effective sample size by studying the effect of phase-correlation on 
the variance of the mean phase estimator. For an order of magnitude estimate, 
we assume the sample consists of $m$ structures of particles in phase space, 
with structure $i$ containing $n_i$ particles. Let us consider the idealised 
case in which the particles in each structure have the same phase-space 
coordinate, with phase angle $\theta_i$ for structure $i$. Now the mean phase 
is 
\begin{equation}
\bar{\theta}=\sum_{i=1}^m w_i \theta_i,
\end{equation}
where $w_i=n_i/\sum n_i$. Its variance is
\begin{equation}
\sigma^2_{\bar{\theta}}=\sum_{i=1}^m w_i^2 \sigma_0^2,
\end{equation} where $\sigma_0^2$ is the variance of a single phase angle 
$\theta_i$. Without phase correlation (i.e., $n_i=1$), we would have 
\begin{equation}
\sigma^2_{\bar{\theta},0}=\frac{1}{N} \sigma_0^2,
\end{equation} where $N=\sum n_i$. So the effective sample size is
\begin{align}
N_{\rm stream,eff}&=N \frac{\sigma^2_{\bar{\theta},0}}{\sigma^2_{\bar{\theta}}}\\
&=\frac{(\sum n_i)^2}{\sum n_i^2}.
\end{align}

\section{Deriving oPDF in action-angle coordinates}\label{sec:equiv_continuity}
Following Han et al. 2015a, we derive the orbital PDF in action-angle coordinates $\{Q_i, 
\theta_i\}$, where $\{Q_i\}$ ($i=1,..3$) are the actions and $\{\theta_i\}$ are the corresponding angles. 
In this coordinate system, the time-independent collisionless Boltzmann equation reads
 \begin{equation}
  \sum_i \left(\frac{\partial(f\dot{Q_i})}{\partial Q_i} + \frac{\partial (f\dot{\theta_i})}{\partial \theta_i}\right)=0.
 \end{equation}
 Since the actions are conserved, i.e., $\dot{Q_i}=0$, we have
 \begin{equation}\label{eq_continuity}
  \sum_i \frac{\partial (f\dot{\theta_i})}{\partial \theta_i}=0.
 \end{equation}
 By the definition of action-angles, $\dot{\theta_i}$ is constant, so that 
 \begin{equation}
  \sum_i \dot{\theta_i}\frac{\partial f}{\partial \theta_i}=0.
 \end{equation}
 This means the derivative of $f$ along the direction of motion, $\dot{\vec{\theta}}$, is zero, 
 that is, $f$ is constant along the orbit. Because the probability density of a particle in $\theta$ space is given by
\begin{equation}
 \frac{d P}{d^3\theta}\big\vert Q \propto f,
\end{equation} we have
\begin{equation}
 d P(\{\theta_i\}|\{Q_i\})\propto d^3\theta
\end{equation} along the orbit.

\end{document}